\documentclass[review]{elsarticle}

\usepackage{lineno,hyperref}
\modulolinenumbers[5]
\usepackage{graphicx, color}

\begin{document}

\begin{frontmatter}

\title{The application of the Lattice Boltzmann method  to the  one-dimensional modeling of pulse waves in elastic vessels}

\author{Oleg  Ilyin}

\address{Dorodnicyn Computing Centre of Russian Academy of Sciences, Vavilova st. 40, 119333 Moscow, Russia}
\ead{oilyin@gmail.com}
\vspace{10pt}

\begin{abstract}
The one-dimensional nonlinear equations for the  blood flow motion in distensible vessels are considered using the kinetic approach. It is shown that the Lattice Boltzmann (LB) model  for non-ideal gas is asymptotically equivalent to the blood flow equations for compliant vessels at the limit of low Knudsen numbers. The equations of  state for  non-ideal gas are transformed to the pressure-luminal area response.  This property allows to model arbitrary pressure-luminal area relations. 
Several test problems are considered: the propagation of a sole  nonlinear wave in an elastic vessel, the  propagation of a pulse wave  in a vessel with varying mechanical properties (artery stiffening) and in an artery bifurcation, in the last problem Resistor-Capacitor-Resistor (RCR) boundary conditions are considered. The comparison with the previous results show a good precision. 
\end{abstract}


\end{frontmatter}

\section{Introduction}

The models of blood motion in  cardiovascular system vary from the $0$D lumped models, $1$D pulse propagation equations to $3$D viscous flow equations \cite{2004taylor}. In many cases the $3$D approach based on the solution of the Navier-Stokes equations is too detailed while $0$D  lumped models are oversimplified and applicable only for the distal vasculature. In the case of the $1$D models it is assumed that the radial velocity is negligible \cite{2003sherwin}. Then, integrating the Navier-Stokes equations over the radial variable the $1$D nonlinear system of equations (depending only on one spatial variable, axial coordinate) for  luminal area change (or  pressure) and  axial blood velocity is derived \cite{1973hughes,2003sherwin}. Several numerical approaches for solving of these equations have been proposed  
including variations of  Galerkin  discretization  \cite{2003sherwin, 2003form_lamp, 2008mynard_nithiarasu},  McCormack and Lax-Wendroff schemes \cite{2003MacCormack, 2007hirsch, 2000olufsen, 2019duanmu}, trapezoidal schemes   \cite{2007hou} and a relatively novel well-balance discretization method \cite{2013muller, 2013muller2}. The comparison results of the numerical methods can be found in
 \cite{2015boileau, 2015wang}. Among the open-source numerical solvers one can mention $0$D framework pyNS \cite{2014manini}, $1$D finite-volume OpenBF solver \cite{2017melis},  Lax-Wendroff method based $1$D solver VaMpy \cite{2017diem},  finite element method based Artery.FE framework \cite{2018Agdestein}.

  The  hydrodynamic equations can  be modeled using the kinetic methods like the Lattice Boltzmann approach \cite{2017kruger,2018succi}. This method describes the motion of particles on a Cartesian spatial lattice (advection part), the collision of the particles  is modeled by assuming that the velocity distribution of the particles tends to a local equilibrium state. 
  The Lattice Bolzmann (LB) method correctly reproduces low-Mach incompressible flows like blood motion and can be used for the modeling of the flow in cardiovascular network.

  The Lattice  Boltzmann simulations of the blood flow dynamics in 2D and 3D vessel geometry have gained some popularity recently  \cite{2002fang, 2006chen, 2011melchionna, 2011bisson, 2013pontrelli, 2014derosis}. The boundary conditions for   LB  models applied to blood flow problems in elastic  tubes have  been proposed in \cite{2002fang}.  Several results are devoted to the estimation of  blood endothelial (wall) shear stress in complicated vessel  networks using LB approach, or using LB models coupled with Monte-Carlo simulations \cite{2006chen, 2011melchionna, 2011bisson, 2013pontrelli, 2019Gounley}.

 In the present paper an alternative way to model 1D blood motion equations based on the Lattice Boltzmann method is proposed. It is shown that there exists an  analogy between the one-dimensional LB model D1Q3 with a virtual force (such approach is popular in modeling of non-ideal gas and  multiphase flows using LB approach)  at the  limit  of  small Knudsen numbers (small times between collisions) and  the  blood flow equations in elastic vessels. More precisely, the LB hydrodynamics for D1Q3 model is equivalent to the 1D blood  motion equations if one changes  the  sound velocity in the LB model by the  pulse propagation velocity and the density by the vessel luminal area. The addition of the fictitious  force  allows to model  arbitrary area-pressure vessel responses. The presented method is relatively simple in implementation.
 
 Several test  problems are considered. The method correctly describes the change in shape of the initial pulse wave while  propagating along a vessel, the analytical solution profile is very close  to the LB modeling results. The second test  problem concerns  the  propagation  of   pulse waves with various amplitudes and duration in a vessel with a prosthesis \cite{2003form_lamp, 2003sherwin1}, in the next problem an artery network (5 vessels, 2 bifurcations) is considered  \cite{2003sherwin}.  Finally, the inclusion of the RCR boundary conditions  for the LB model is discussed and LB simulation results are compared with McCormack difference scheme  for the vessel branching supplemented with RCR boundary conditions at the distal ends of the daughter vessels.

\section{Lattice Boltzmann approach for non-ideal gas and 1D blood flow equations in elastic vessels}

\subsection{The mass  and  momentum conservation equations for inviscid 1D blood flow  in elastic vessels} 

 Consider a long elastic vessel filled  with incompressible fluid (blood). The  conservation equations  of mass and momentum in the elastic vessel  are obtained  from  the Navier–Stokes equations 
by the integration over the cross-sectional spatial coordinate 
(radial variable). Assume that the radial component of the blood velocity is close  to zero, and the velocity profile   is flat. 
Then the following equations are obtained \cite{2003sherwin}
\begin{equation}\label{eq01}
\frac{\partial A}{\partial t}+\frac{\partial  Au}{\partial x}=0,
\quad
\frac{\partial u}{\partial t}+\frac{\partial u^2/2}{\partial x}=-\frac{1}{\rho_0}\frac{\partial p}{\partial x}+\frac{\nu}{A}\frac{\partial^2 Au}{\partial  x^2}+f_{dr},
\end{equation}
where $\rho_0$  is the constant blood density (incompressible fluid), $A,u$  is the luminal area, blood velocity respectively and $f_{dr}$  is the viscous drag force.
Here $x$  is the axial coordinate and $p$  is the  blood pressure. 

It is  obvious that one has three unknowns $u,A,p$  and  only the two governing equations. To close  the system  (\ref{eq01})  one needs to introduce a {\it pressure-area relation}
\begin{equation}\label{eq02}
p=f(A),
\end{equation}
where $f$ is a some function, its form should be defined from the elastic  properties of the considered vessel. For realistic vessels this dependence can be  complicated and saturation effects are observed  \cite{1984Lang}. In practice  the Laplace  law  is  popular \cite{2004taylor, 2003sherwin}.

From the  pressure-area relation one can find  the  pulse-wave  velocity using the formula
\begin{equation}\label{pulse_definition}
c_{pulse}^2(A)=\frac{A}{\rho_0}\frac{\partial p}{\partial A}.
\end{equation}
Using the pulse wave velocity definition Eqs (\ref{eq01})-(\ref{eq02}) can be rewritten in the following equivalent form
\begin{equation}\label{eq03}
\frac{\partial A}{\partial t}+\frac{\partial  Au}{\partial x}=0,
\quad
\frac{\partial u}{\partial t}+\frac{\partial u^2/2}{\partial x}=-\frac{c_{pulse}^2(A)}{A}\frac{\partial A}{\partial x}+\frac{\nu}{A}\frac{\partial^2 Au}{\partial  x^2}+f_{dr}.
\end{equation}
In the  present paper the drag force is  not  considered, therefore $f_{dr}=0$.

\subsection{ The  Lattice Boltzmann method  and  hydrodynamics}

It will be  sufficient  to consider the simplest three-velocity model $D1Q3$  \cite{2017kruger}. This model describes three populations of particles traveling on a 1D lattice with equal spacing $\Delta x=c \Delta t$ between the lattice nodes. 
Here $c$ is the  lattice velocity, $\Delta t$ is the lattice  time step. During the discrete time step $\Delta t$ the particles at a node $x$   can hop to adjacent nodes $x \pm c \Delta t$ or stay at the node $x$. The   concentrations of the particles traveling with the velocities $\pm c, 0$  are given by $f_{\pm 1}(t,x)$ and $f_{0}(t,x)$ respectively. The dynamics of $f_{\pm 1}, f_{0}$ obeys the following discrete equations
in space and time
$$
f_{-1}(t+\Delta t, x-c\Delta t)-f_{-1}(t,x)=\frac{\Delta t}{\tau+\frac{\Delta t}{2}}\left(f^{eq}_{-1}(t,x)-f_{-1}(t,x) \right)+\frac{\tau\Delta t }{\tau+\frac{\Delta t}{2}}F_{-1}(t,x),
$$
$$
f_{0}(t+\Delta t, x)-f_{0}(t,x)=\frac{\Delta t}{\tau+\frac{\Delta t}{2}}\left(f^{eq}_{0}(t,x)-f_{0}(t,x) \right)+\frac{\tau\Delta t }{\tau+\frac{\Delta t}{2}}F_{0}(t,x),
$$
$$
f_{1}(t+\Delta t, x+c\Delta t)-f_{1}(t,x)=\frac{\Delta t}{\tau+\frac{\Delta t}{2}}\left(f^{eq}_{1}(t,x)-f_{1}(t,x) \right)+\frac{\tau\Delta t }{\tau+\frac{\Delta t}{2}}F_{1}(t,x),
$$
where $\tau$  is the  relaxation time which can be considered as  a free parameter, $F_{i}$ are the components of an external force  and $f^{eq}_{\pm 1}, f^{eq}_{0}$ are the equilibrium states (analogs  of the Maxwell distribution)
$$
f^{eq}_{-1}(t,x)=w_{-1}\rho(t,x)\left(1-3\frac{u(t,x)}{c}+3\frac{u(t,x)^2}{c^2}\right),
$$
$$
f^{eq}_{0}(t,x)=w_{0}\rho(t,x)\left(1-3\frac{u(t,x)^2}{2c^2}\right),
$$
$$
f^{eq}_{1}(t,x)=w_{1}\rho(t,x)\left(1+3\frac{u(t,x)}{c}+3\frac{u(t,x)^2}{c^2}\right),
$$
where $w_{\pm 1}=1/6, w_0=4/6$ are the lattice weights and $\rho, u$ are the density and flow velocity
$$
\rho(t,x)=f_{-1}(t,x)+f_0(t,x)+f_{+1}(t,x),
$$
$$
\rho(t,x) u(t,x)=(-f_{-1}(t,x)c+f_{+1}(t,x)c)+\frac{\Delta t}{2}\sum_i F_i(t,x)c_i,
$$
moreover, one can define the  full energy of the  lattice gas in every node  by the formula
$$
\rho(t,x)(u(t,x)^2+c_s^2)=(f_{-1}(t,x)+f_{+1}(t,x))c
$$
and here $c_s$ is the sound  velocity of the lattice gas defined  by
$$
\quad c_s=\sqrt{\frac{1}{3}}c.
$$
Here the  external force  is  taken in the  linearized form \cite{2017kruger}
$$
F_i=w_{i}(c_i/c_s^2)a, \quad  i=\pm 1, 0,
$$
where $a$ is the  force  magnitude.

At the limit of small values  of $\tau$ the lattice  gas  can be considered as  continuous media since the time between collisions tends  to zero ( $\tau$  is  proportional to the expected time between particle  collisions). In the continuous limit the considered LB model describes some hydrodynamics which can be  obtained  using  the Chapman-Enskog expansion on a small parameter, the detailed analysis of this procedure for LB method in 1D case can be found in \cite{2007karlin}. One has
\begin{equation}\label{lb_hydr}
\frac{\partial\rho}{\partial t}+\frac{\partial \rho u}{\partial x}=0,
\quad \frac{\partial u}{\partial t}+\frac{\partial u^2/2}{\partial x}=-\frac{c_s^2}{\rho}\frac{\partial \rho}{\partial x}
+\frac{2\nu}{\rho}\frac{\partial}{\partial x}\left(\rho \frac{\partial u}{\partial x}\right) +\frac{a}{\rho}+O(u^3)+O(\Delta t^2),
\end{equation}
where $O(u^3)$ is the spurious term which appears due to the  quadratic form  of the local equilibrium on the bulk velocity $u$, $O(\Delta t^2)$ is space-time discretization error ( LB scheme has  second order accuracy) and $\nu=c_s^2\tau$ is the gas viscosity. If the  flow  is slow $u << c_s$ (i.e the Mach number  $Ma=u/c_s$ is small) and the time step is small then one can assert that the LB model correctly describes 1D  hydrodynamics. It should be  mentioned  that the Navier-Stokes equations   without spurious $u^3$ term can be  obtained if high-order LB models are considered \cite{2006chikatamarla, 2009chikatamarla}. In the  present case the term proportional $u^3$ is  unimportant since  the blood flow is significantly subsonic.

\subsection{ The  Lattice Boltzmann equation as a model of hemodynamics}
One can see that the luminal area $A$ in the equations (\ref{eq03}) is very similar   to the density $\rho$ in the equations (\ref{lb_hydr}).
If one changes the density $\rho$ by the luminal area $A$   and the sound velocity $c_s$  by the pulse wave velocity  $c_{pulse}$ then Eqs (\ref{lb_hydr}) turn into  Eqs (\ref{eq03})  if the  flow  is slow, discretization error is  negligible.  These conditions are realistic due to a) for a typical blood flow the Mach number is over $0.1$,  $u << c_{pulse}$ \cite{2003sherwin}    b) the discretization error can be  controlled by the choice of $\Delta t$.

The lattice Boltzmann approach gives slightly different longitudinal viscosity term $ \frac{\partial}{\partial x}\left(A \frac{\partial u}{\partial x}\right)$ than $\frac{\partial}{\partial x}\left(\frac{\partial Au}{\partial x}\right)$ in Eqs (\ref{eq01})-(\ref{eq02}) . This difference  is unimportant since the terms are very close (same in the linear theory), moreover stream-wise diffusion  effects  play
 small role in hemodynamics and usually omitted from the  consideration. In the  present case  the diffusion is  an inherent  property of the LB method.

One should emphasize  that the force  free  case in the equations (\ref{lb_hydr})  describes only the particular case  of constant pulse velocity $c_{pulse}$ since the sound velocity $c_s$ is constant, the blood flow equations are derived for the  pressure-area relation in the form $p \sim \log(A)$ (this results from the definition of the pulse wave velocity (\ref{pulse_definition})). The  logarithmic dependence  is  only qualitatively correct, a more  popular  Laplace  law states that $p \sim \sqrt{A}$, moreover the dependence  of the  pressure on the luminal  area  in real arteries is saturating \cite{1984Lang},  hence  a  generalization of the model is desirable.  

 The  idea of the generalization of the presented approach  is  to choose the  force term  in   such a way that the case  of the  non-constant  pulse  velocity  is captured. This approach is  similar to the method of modeling  of  non-ideal gases and  phase transitions  using LB schemes  \cite{2003zhang, 2006yuan, 2009kuper, 2010kuper, 2012li, 2018kuper}.  Following the articles \cite{ 2009kuper, 2010kuper, 2018kuper}  the force  is  introduced  using a  pseudo-potential $U$
 \begin{equation}\label{force_definition}
    F= -w_{i}(c_i/c_s^2)\frac{\partial U}{\partial x}, \quad U=h(\rho)-\rho c_s^2,
 \end{equation}
where $h(\rho)$ is a some function. Using the  expressions (\ref{force_definition})  the terms  $-\frac{c_s^2}{\rho}\frac{\partial \rho}{\partial x}+ \frac{a}{\rho}$    become  $-\frac{1}{\rho}\frac{\partial h}{\partial \rho}\frac{\partial \rho}{\partial x}$ in the   hydrodynamic  equations (\ref{lb_hydr}), next
changing $\rho$ by $A$  and  selecting $h(A)$ in a such  way that 
$$
\frac{\partial h(A)}{\partial A}=c_{pulse}^2(A)=\frac{A}{\rho_0}\frac{\partial p(A)}{\partial A}
$$
where $p(A)$ is the  desired pressure-area law  one obtains the equations (\ref{eq03}).

Following  \cite{2010kuper} the  pseudo-potential $U$ can be written in the form $U=-\Phi^2$, $\Phi=\sqrt{Ac_s^2-h(A)}$ and  for the  LB  model the  central  discretization of the  spatial derivatives is adopted
\begin{equation}\label{force_discretize}
\frac{\partial U}{\partial x}=-2 \Phi \frac{\partial \Phi}{\partial x} \approx -2\Phi(x)\frac{\Phi(x+\Delta x)-\Phi(x-\Delta x)}{2\Delta x},
\end{equation}
where $\Delta x =c \Delta t$ is  the lattice spatial step. For the LB D1Q3 model with the force discretization in the  form  (\ref{force_discretize}) the stability conditions are    known \cite{2010kuper} 
\begin{equation}\label{stability}
\frac{\partial h(A)}{\partial A} \leq c^2+c_s^2= \frac{4}{3}c^2.
\end{equation}
From the relation (\ref{stability}) one can see that stability can  be  controlled by adjusting  the  lattice velocity $c=\Delta x/\Delta t$.  

The model implementation at the each time step consists of the  two parts. Firstly, the collision term (right hand side  of LB model) is computed in all spatial nodes   and then the post-collision distribution functions $f_i^*(t,x)$ are obtained. Next, the streaming step is performed $f(t+\Delta t, x+c_i\Delta t)=f_i^*(t,x)$.

\begin{figure}
\centering
\includegraphics[width=0.8\textwidth]{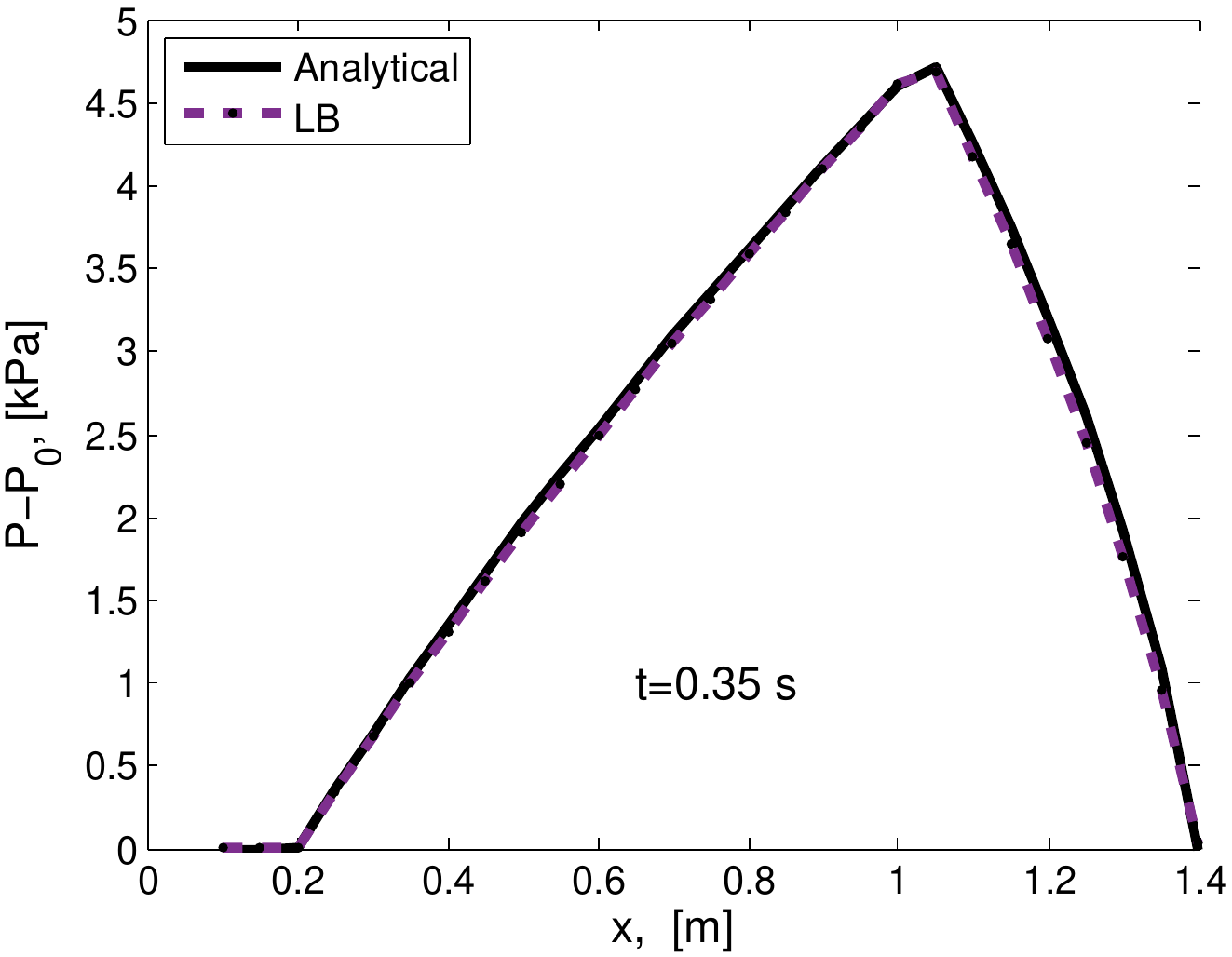}
\caption{ The pulse wave snapshot $p-p_0$  at $t=0.35 \, s$. 
The presented profile at  $t=0.35$ is skewed  due  to nonlinear effects, the viscous terms do not affect wave form. 
The initial symmetrical triangle-shaped pulse
wave has the physiological duration of $t_0=0.3 \, s$, the blood density $\rho_0=10^3 kg/m^3$, the pulse wave  velocity in the linearized blood  flow equations is defined as $c_{pulse}(A_0)=\sqrt{A_0^{1/2}/\rho_0 D_0}$ and equals $4 m/s$, the unperturbed luminal area $A_0$ equals $ 7\times 10^{-4} m^2 $. The  constant  $a$ is  taken such that the  initial maximal relative luminal area  change is $1.2$   i.e. $at_0=1.2 A_0$.
The  relaxation time $\tau$ satisfies  the  condition $\nu=c_s^2\tau=4 \times 10^{-6} \, m^2/s$, where $\nu$ is  the  blood viscosity.
 }
 \end{figure}

\begin{figure}
\centering
\includegraphics[width=0.95\textwidth]{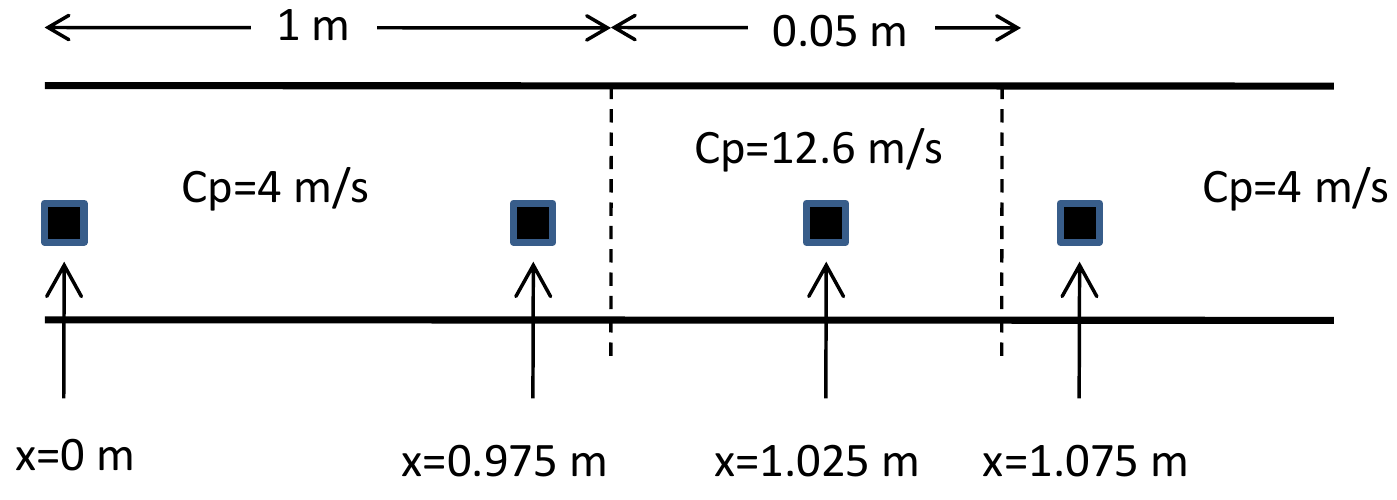}
\caption{ Modeled vessel geometry (vessel with a stent or prosthesis).
The  stented  region ( stiff domain) is placed between $x=1 \, m$ and $x=1.05 \, m$. The pressure waveforms are measured at $x=0 \, m, \, x=0.975 \, m, x=1.025 \, m$. The  boundaries  of the stiff region  are denoted  by dashed lines. $Cp$ denotes the value of the  linearized  pulse velocity $c_{pulse}(A_0)$.  }
 \end{figure}

\begin{figure}
   \centering
   \begin{minipage}[t]{.99\textwidth}
       \includegraphics[width=0.98\textwidth]{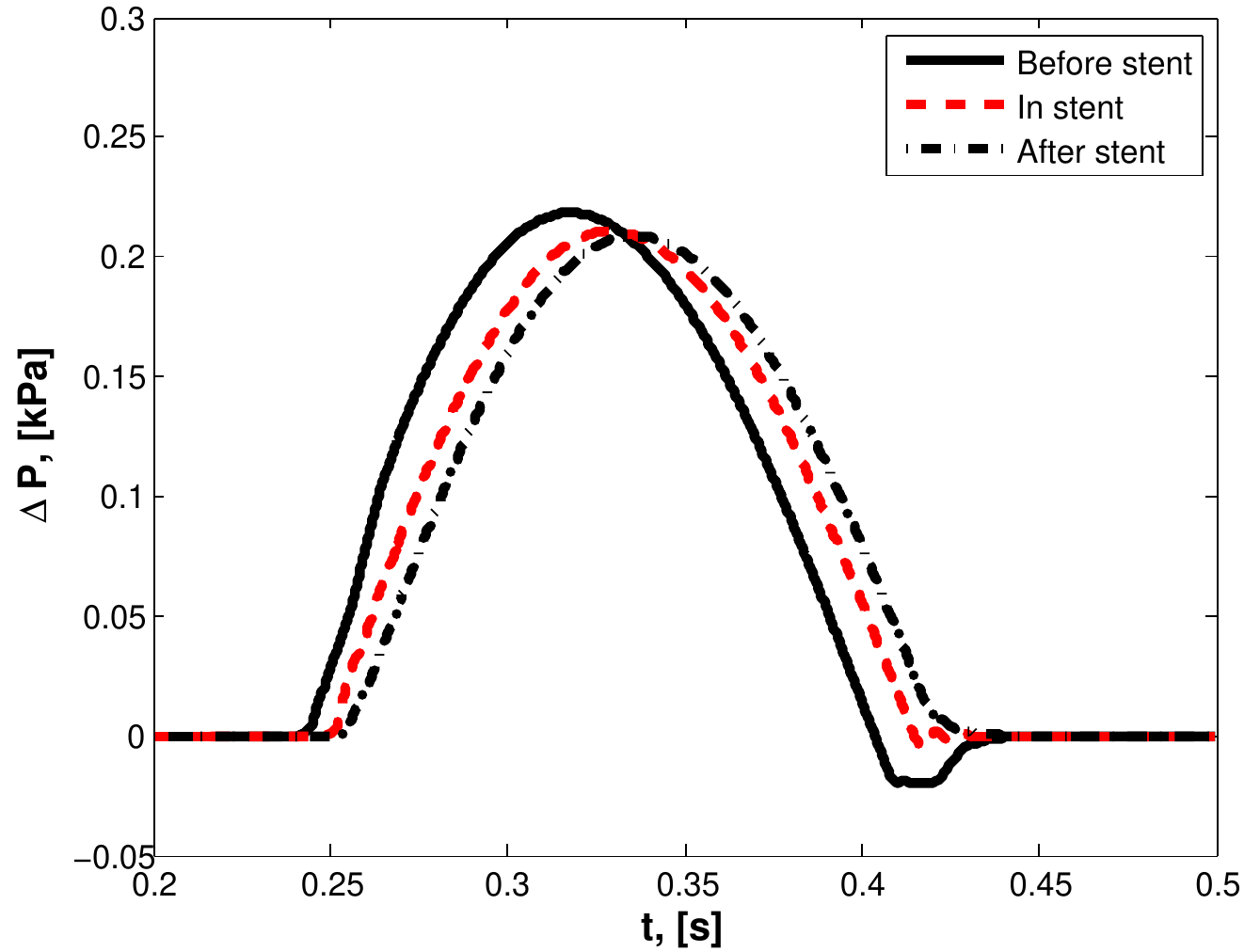}
   \end{minipage}
   \caption{ 
   The pulse wave  profiles $p-p_0$ at
   $x=0 m$ (initial sinusoidal wave, left  slide) and $x=0.975 \, m$ (before stented region), $x=1.025 \, m$ (in the middle of the stented region), $x=1.075 \, m$ (after stent). The initial sinusoidal pressure pulse wave has the maximal amplitude of $0.2\, kPa$ and the duration of $0.165 \, s$.  }
\end{figure}

\begin{figure}
\centering
\includegraphics[width=0.95\textwidth]{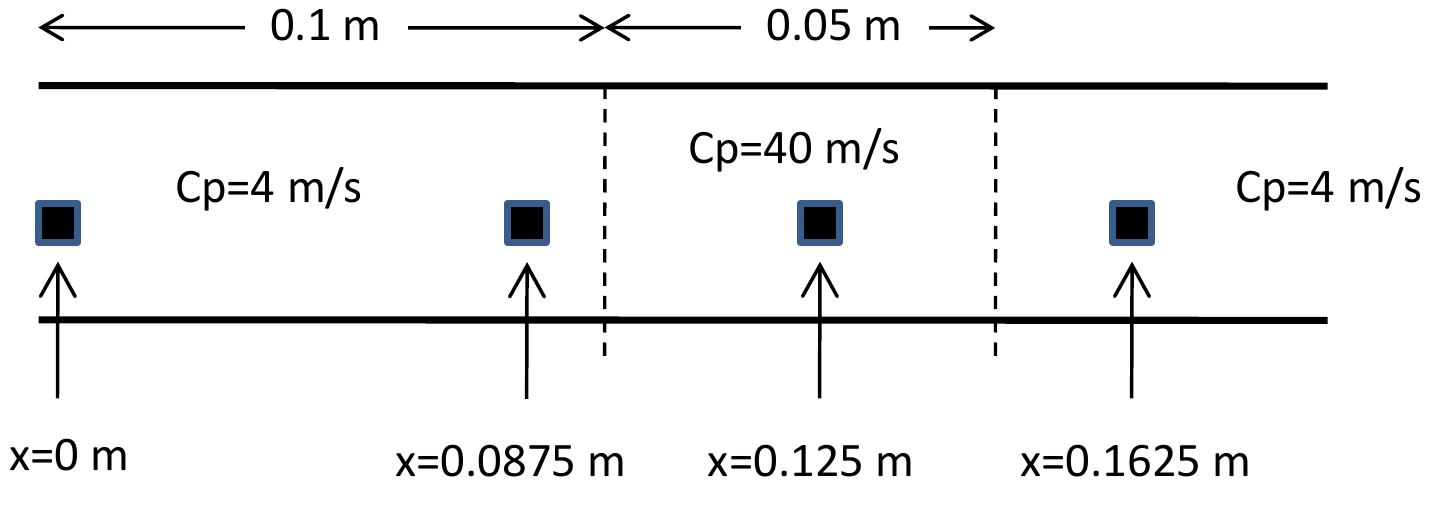}
\caption{ Modeled vessel geometry (vessel with a stent or prosthesis) for  a short pressure impulse.
The  stented  region (stiff domain) is placed between $x=0.1 \, m$ and $x=0.15 \, m$. The pressure waveforms are measured at $x=0 \, m, \, x=0.0875 \, m, x=0.125 \, m, x=0.1625 \, m$. The  boundaries  of the stiff region  are denoted  by dashed lines. $Cp$ denotes the value of the  linearized  pulse velocity $c_{pulse}(A_0)$. }
 \end{figure}

\begin{figure}
   \centering
  \begin{minipage}[t]{.99\textwidth}
       \includegraphics[width=0.95\textwidth]{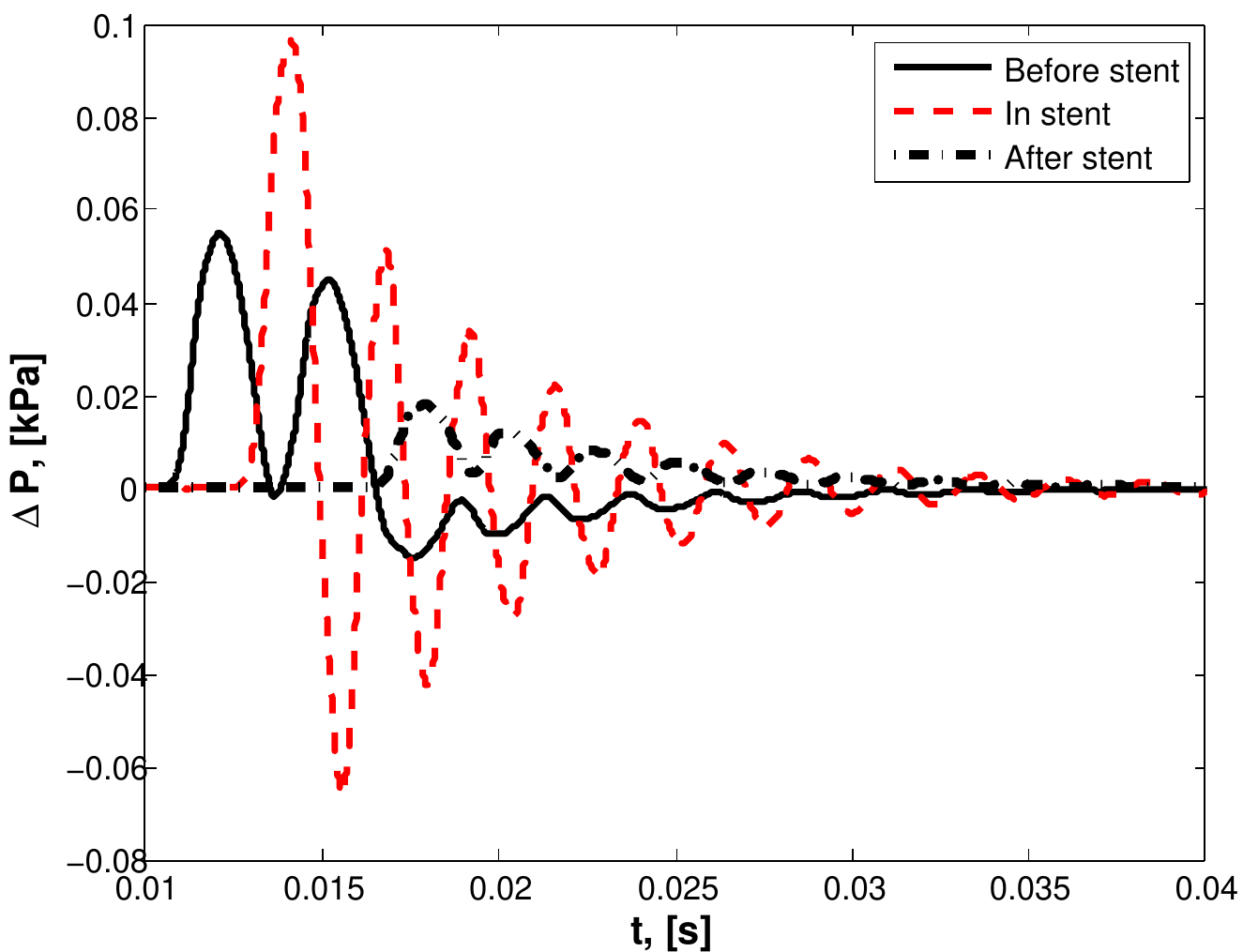}
 \end{minipage}
   \caption{ 
   The pulse wave  profiles $p-p_0$ at
   $x=0 \, m$ (short sinusoidal impulse, left  slide) and $x=0.0875 \,$ (before stented region), $x=1.125 \, m$ (in the middle of the stented region), $x=1.1625 \, m$ (after stent). The  initial  pressure  impulse has the amplitude of $0.055 \, kPa$ and  the  duration of $0.0025 \, s$.  }
\end{figure}

\subsection{Stability improvement}
For the  conventional LB method   Courant-Friedrichs-Lewy (CFL) number (defined as $c \Delta t/\Delta x$ ) equals to unity. This  means  that LB method  is only marginally stable. This feature  is well-known and several efforts have  been paid  to improve the stability properties, the possible solution is to rewrite the streaming step in such a form that LB model becomes a finite-difference scheme with controllable CFL number   \cite{1997Qian, 2001zhang, 2001guo, 2003sofonea, 2004sofonea, 2018guo}. 

The simplest way to decrease CFL number is the implementation of the 
fractional time step.
In this  method \cite{2001zhang} the  streaming part $f_i(t+\Delta t, x+c_i \Delta t)=f_i^*(t, x)$, where $c_i=(-c,0,c)$ and $f_i^*$ is the  post-collision distribution function  is  replaced  by
$$
f_i(t+\Delta t, x+c_i \Delta t)=CFL\left(f_i^*(t,x)+\frac{(1-CFL)}{2}\delta f_i^*(t,x)\right)+
$$
\begin{equation}\label{fractstep}
+(1-CFL)\left(f_i^*(t,x+c_i\Delta t)-\frac{CFL}{2}\delta f_i^*(t,x+c_i\Delta t)   \right),
\end{equation}
where  $\delta f_i^*(t,x)=f_i^*(t, x+c_i \Delta t)-f_i^*(t, x)$. For this realization of the streaming component the particles travel
a distance $CFL c \Delta t$  during a time step $\Delta t$ and the parameter $0<CFL \leq 0$ is Courant-Friedrichs-Lewy number. The  form  (\ref{fractstep}) guarantees second order accuracy in space and  time. Finally, one should mention that for the LB model with the streaming step  (\ref{fractstep}) the viscosity is governed  by the relation $\nu=CFL c_s^2\tau$ \cite{2001zhang}.

In the  present  paper  for the  test problems in Paragraphs 3.1-3.3 the conventional streaming part ($CFL=1$) was applied, for the test  problem in Paragraph 3.4 the fractional time step with $CFL=0.5$ was implemented.


\section{The  model application.}

From the  results  of the  previous  paragraph one derives  the  following  LB model
$$
f_{-1}(t+\Delta t, x-c\Delta t)-f_{-1}(t,x)=\frac{\Delta t}{\tau+\frac{\Delta t}{2}}\left(f^{eq}_{-1}(t,x)-f_{-1}(t,x) \right)-
$$
\begin{equation}\label{fin_eq01}
-\frac{\tau}{2c^2\left(\tau+\frac{\Delta t}{2}\right)}\Phi(x)\bigl(\Phi(x+c\Delta t)-\Phi(x-c\Delta t)\bigr),
\end{equation}
\begin{equation}\label{fin_eq02}
f_{0}(t+\Delta t, x)-f_{0}(t,x)=\frac{\Delta t}{\tau+\frac{\Delta t}{2}}\left(f^{eq}_{0}(t,x)-f_{0}(t,x) \right),
\end{equation}
$$
f_{1}(t+\Delta t, x+c\Delta t)-f_{+1}(t,x)=\frac{\Delta t}{\tau+\frac{\Delta t}{2}}\left(f^{eq}_{1}(t,x)-f_{1}(t,x) \right)+
$$
\begin{equation}\label{fin_eq03}
+\frac{\tau}{2c^2\left(\tau+\frac{\Delta t}{2}\right)}\Phi(x)\bigl(\Phi(x+c\Delta t)-\Phi(x-c\Delta t)\bigr),
\end{equation}
where  $\Phi(x)=\sqrt{Ac_s^2-h(A)}$ and $c_s^2=(c^2/3)$, $c =\Delta x/\Delta t$ . The function  $h(A)$ is obtained  from the equation
$$
\frac{\partial h(A)}{\partial A}=c_{pulse}^2(A)=\frac{A}{\rho_0}\frac{\partial p(A)}{\partial A},
$$  
here  $c_{pulse}(A), p(A)$ are  the target  pressure pulse  velocity and  pressure-area relation.  The local equilibrium states $f^{eq}_{\pm 1}, f^{eq}_{0}$ are defined as
$$
f^{eq}_{\pm 1}=\frac{A}{6}\left(1\pm 3\frac{u}{c}+3\frac{u^2}{c^2}\right),
\quad f^{eq}_{0}=\frac{4A}{6}\left(1-3\frac{u^2}{2c^2}\right),
$$
where   
$$
A(t,x)=(f_{-1}+f_{0}+f_{1})(t,x),
$$
$$
A(t,x)u(t,x)=(f_1(t,x)-f_{-1}(t,x))c+\frac{c}{6c_s^2}\Phi(t,x)(\Phi(t,x+c\Delta t)-\Phi(t,x-c\Delta t)).
$$
 The  relaxation time $\tau$ should  be taken small enough if the inviscid  flow is  modeled. Note  that wall friction forces are  not taken in account but potentially  they can  be implemented  as an additional  external  force.
The  pulse  pressure  waves presented in figures are  recovered with use  of the  pressure-area  relation $p(A)$.

The case  of LB model without external force corresponds to the following pressure-area relation
\begin{equation}\label{p3exp_log}
p=p_0+\frac{1}{D_0}\log\left(\frac{A}{A_0}\right).
\end{equation}
For the hemodynamic modeling the  following tube law is popular
\begin{equation}\label{p3exp_p}
p=p_0+\frac{1}{n D_0}(A^{n}-A_0^{n}),
\end{equation}
where $D_0, p_0,\rho_0$ are the   vessel distensibility, blood pressure at rest,  blood density and $n>0$. The case $n=1/2$ corresponds  to the Laplace  law.
The value of  $p_0$ can  be  taken  arbitrary since only $p-p_0$   is computed. It will be more convenient to use a linearized  pressure pulse  velocity $c_{pulse}(A_0)$ instead of  the  distensibility $D_0$  
$$
c_{pulse}(A_0)=\sqrt{A_0^n/\rho_0 D_0}.
$$

\subsection{A sole forward pulse wave  propagation.}
The evolution of  the   nonlinear wave  which  has  triangle  shaped  form  in the  initial moment  of  time  is considered in a  semi-infinite vessel $x>0$ with constant elastic properties.  

For a forward traveling wave one has the relation  between the wave velocity $u$ and the luminal area $A$ \cite{2018ilyin}
\begin{equation}\label{06subst}
u=\int_{A_0}^{A}\frac{dz}{\sqrt{\rho_0 D_0/z^{n}}z},
\end{equation}
then the equations (\ref{eq01})-(\ref{eq02}) reduce to a differential equation for $A$
\begin{equation}\label{05onedeq1}
\frac{\partial A}{\partial t}+\frac{1}{\sqrt{\rho_0 D_0}}\left (\left(1+\frac{2}{n}\right)A^{n/2}-\frac{2}{n}A_0^{n/2}  \right )\frac{\partial A}{\partial x}=0.
\end{equation}
The equation (\ref{05onedeq1}) is  supplemented  with the symmetric triangle-shaped initial-boundary  condition  at $x=0$ for $t \in [0,T_0]$, where $T_0$ is  one  heartbeat
\begin{equation}\label{05bounds1}
A(t,x)|_{x=0}=A_0+at,\quad t\in [0, t_0),
\end{equation}
\begin{equation}\label{05bounds2}
A(t,x)|_{x=0}=A_0+a(2t_0-t), \quad t \in [t_0, 2t_0],
\end{equation}
\begin{equation}\label{05bounds3}
A(t,x)|_{x=0}=A_0,\quad t \in (2t_0,T_0]
\end{equation}
and
\begin{equation}\label{05bounds4}
A(t,x)|_{t=0}=A_0,\quad x>0.
\end{equation}

The pressure-area  relation (\ref{p3exp_p}),$n=1/2$, which corresponds to the Laplace law is considered. The analytical solutions to this  problem are  obtained in \cite{2018ilyin}. For the sake of brevity this solution is not given here.
The LB equations (\ref{fin_eq01})-(\ref{fin_eq03}) are applied with the initial-boundary conditions 
(\ref{05bounds1})-(\ref{05bounds4}) and the modeling results are  compared   with the analytical solutions at  $t=0.35 \, s$. The  modeling constants are  given in Fig. 1 caption. A  very good agreement between the solutions is observed ( Fig. 1).

\begin{figure}
   \centering
  \begin{minipage}[t]{.99\textwidth}
       \includegraphics[width=0.95\textwidth]{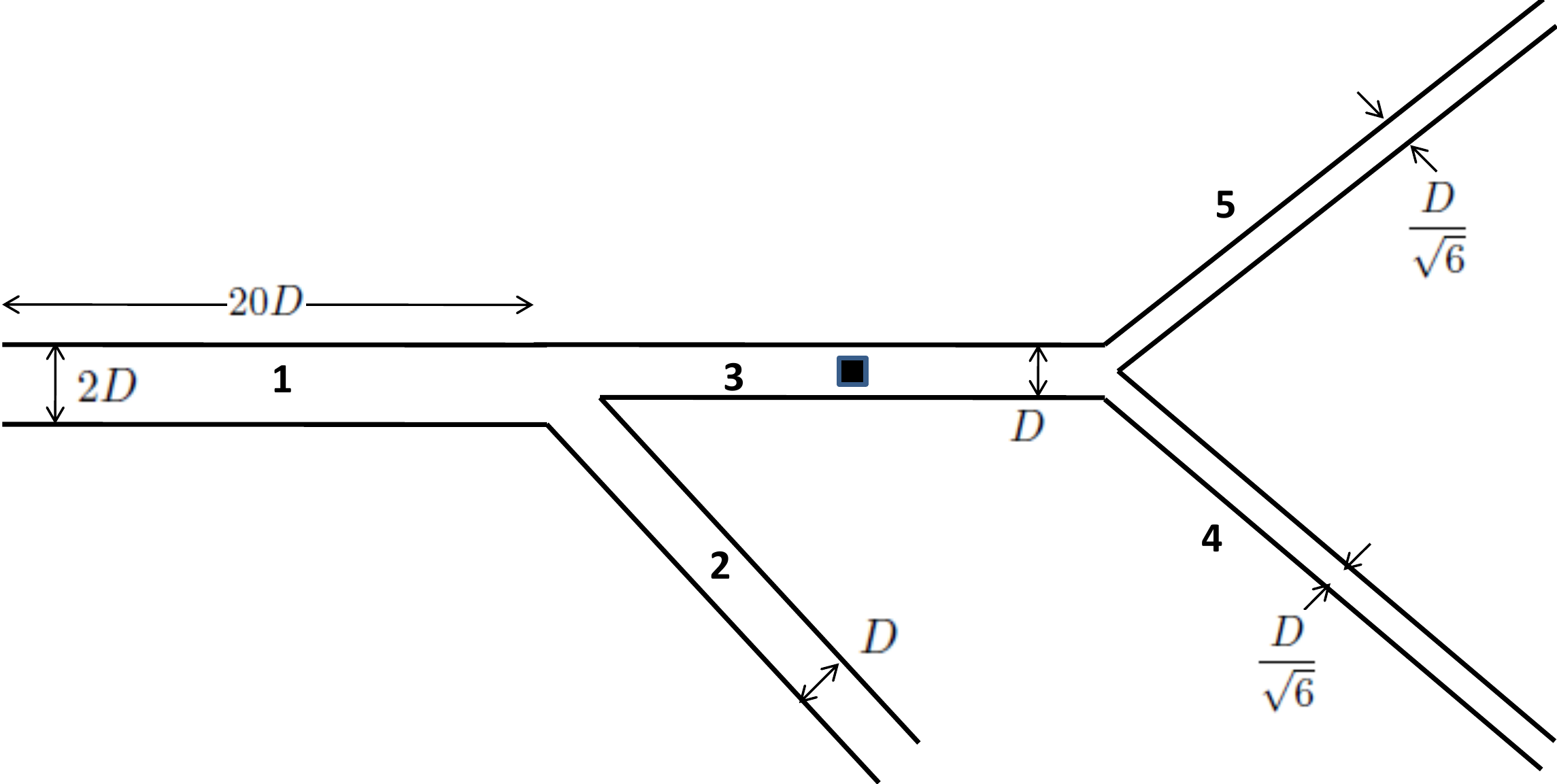}
 \end{minipage}
   \caption{ Modeled vessel branching geometry. All vessels  have the same length of $20D$, where $D=2.5 \times 10^{-2} \, m$. The wave  is  measured at the  middle of the third vessel (black square).
     }
\end{figure}
\begin{figure}
   \centering
  \begin{minipage}[t]{.99\textwidth}
       \includegraphics[width=0.48\textwidth]{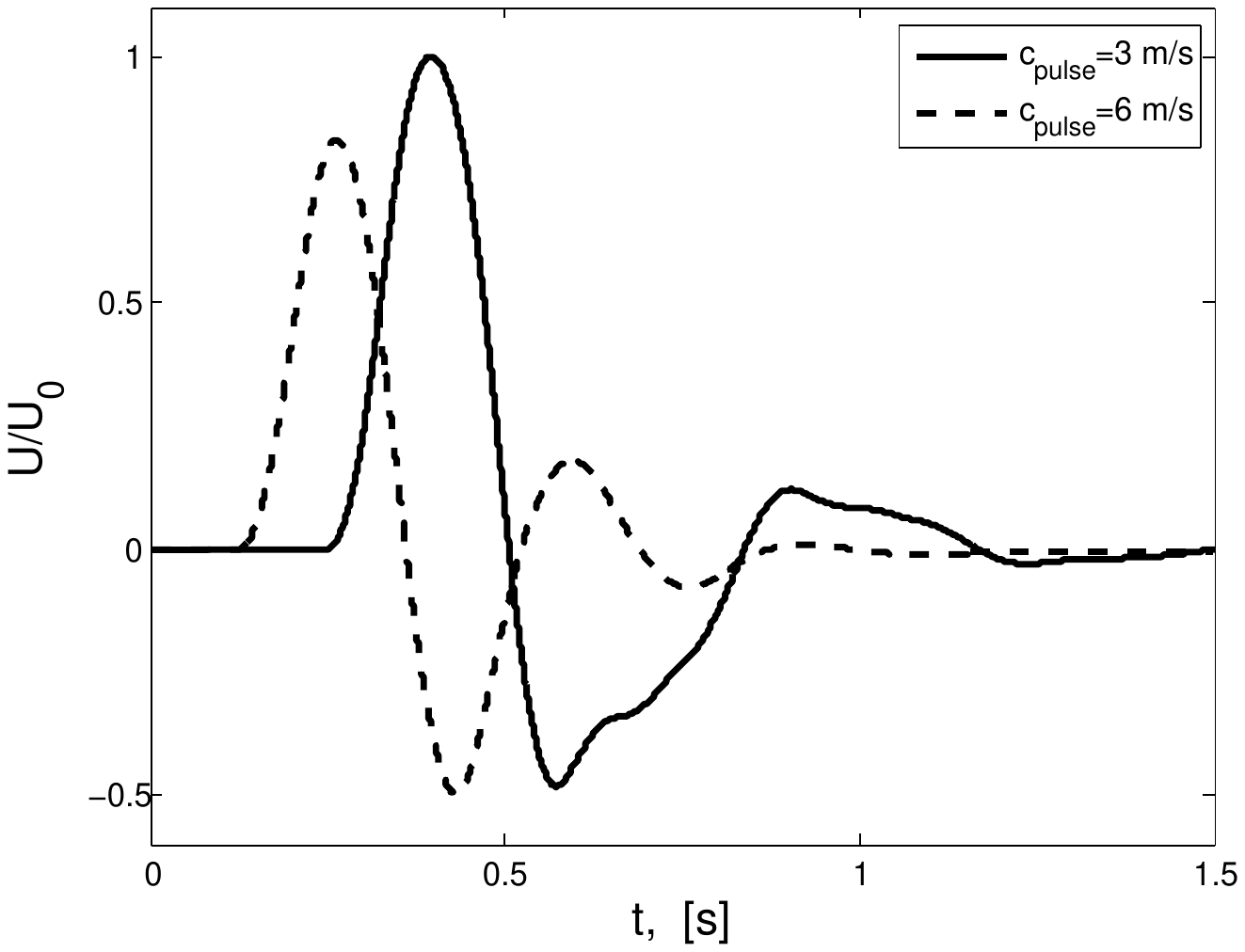}
       \includegraphics[width=0.48\textwidth]{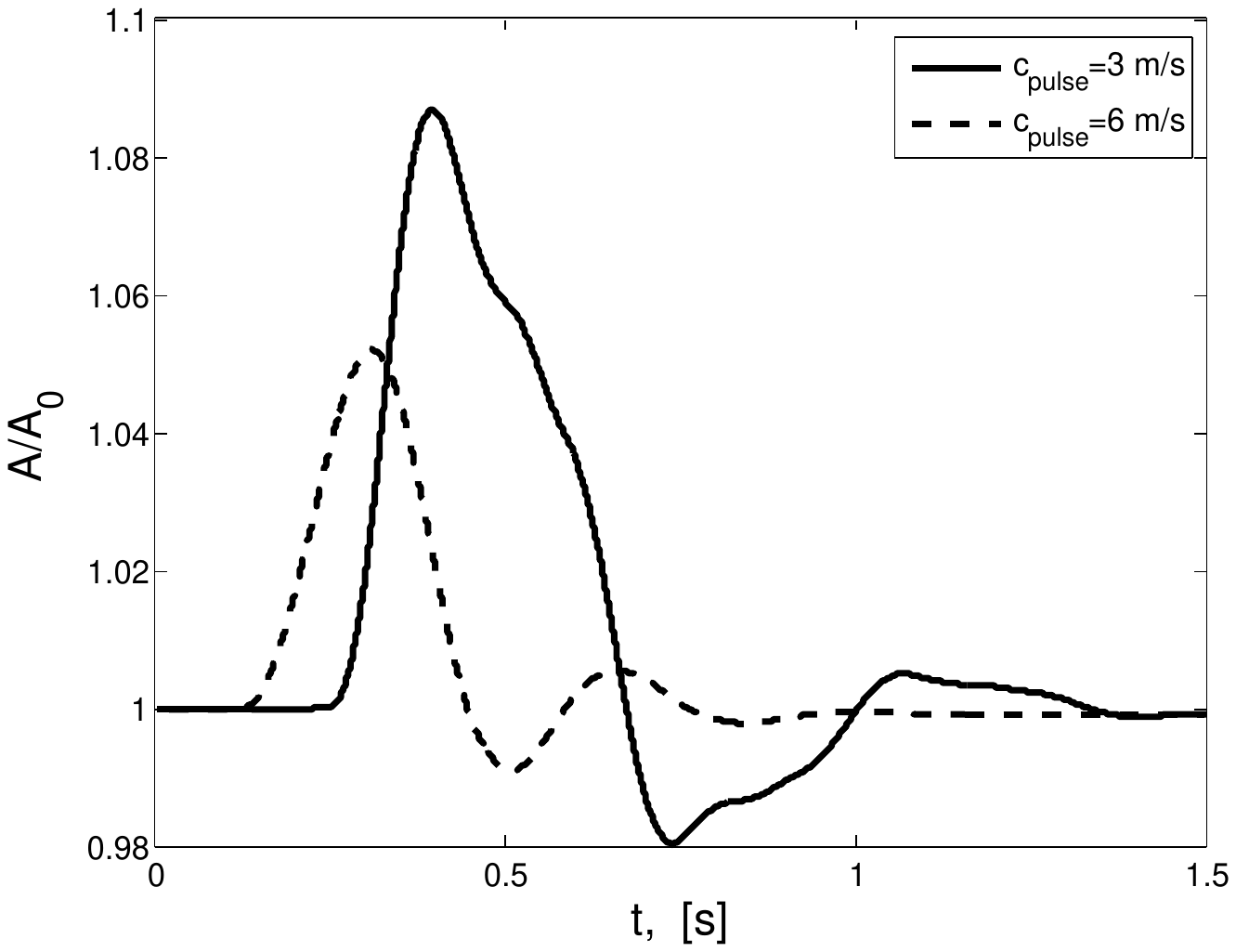}
 \end{minipage}
   \caption{The blood flow velocity (normalized at the initial flow velocity amplitude $u_0=0.25 \, m/s$) and luminal area  profiles (normalized at the undisturbed vessel area $A_0^{(3)}$) measured at the middle of the vessel $3$. Solid lines  correspond  to the linearized pulse wave velocity of $3 \, m/s$ , the dashed for $6 \, m/s$.}
\end{figure}
\begin{figure}
   \centering
  \begin{minipage}[t]{.99\textwidth}
       \includegraphics[width=0.48\textwidth]{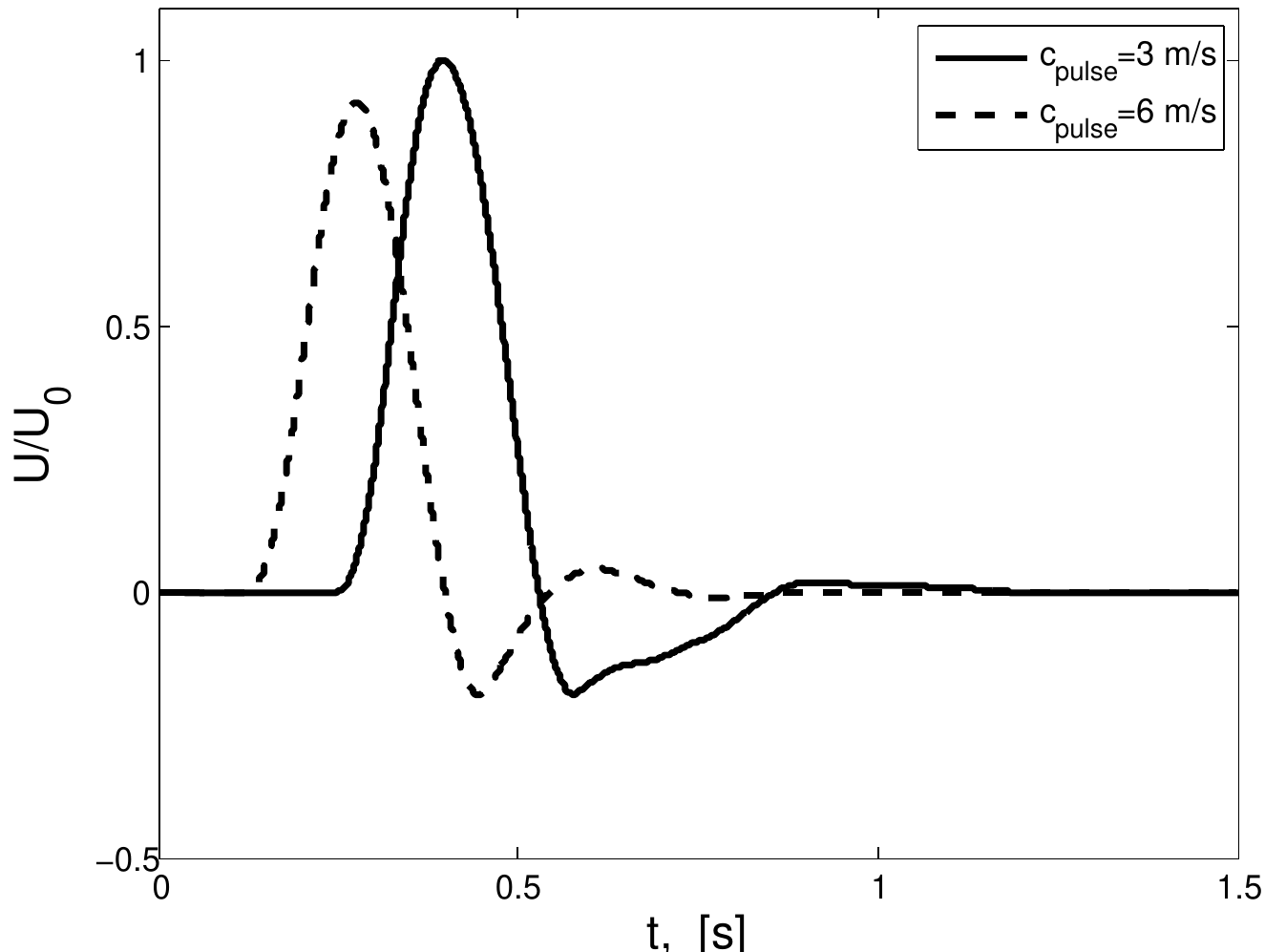}
       \includegraphics[width=0.48\textwidth]{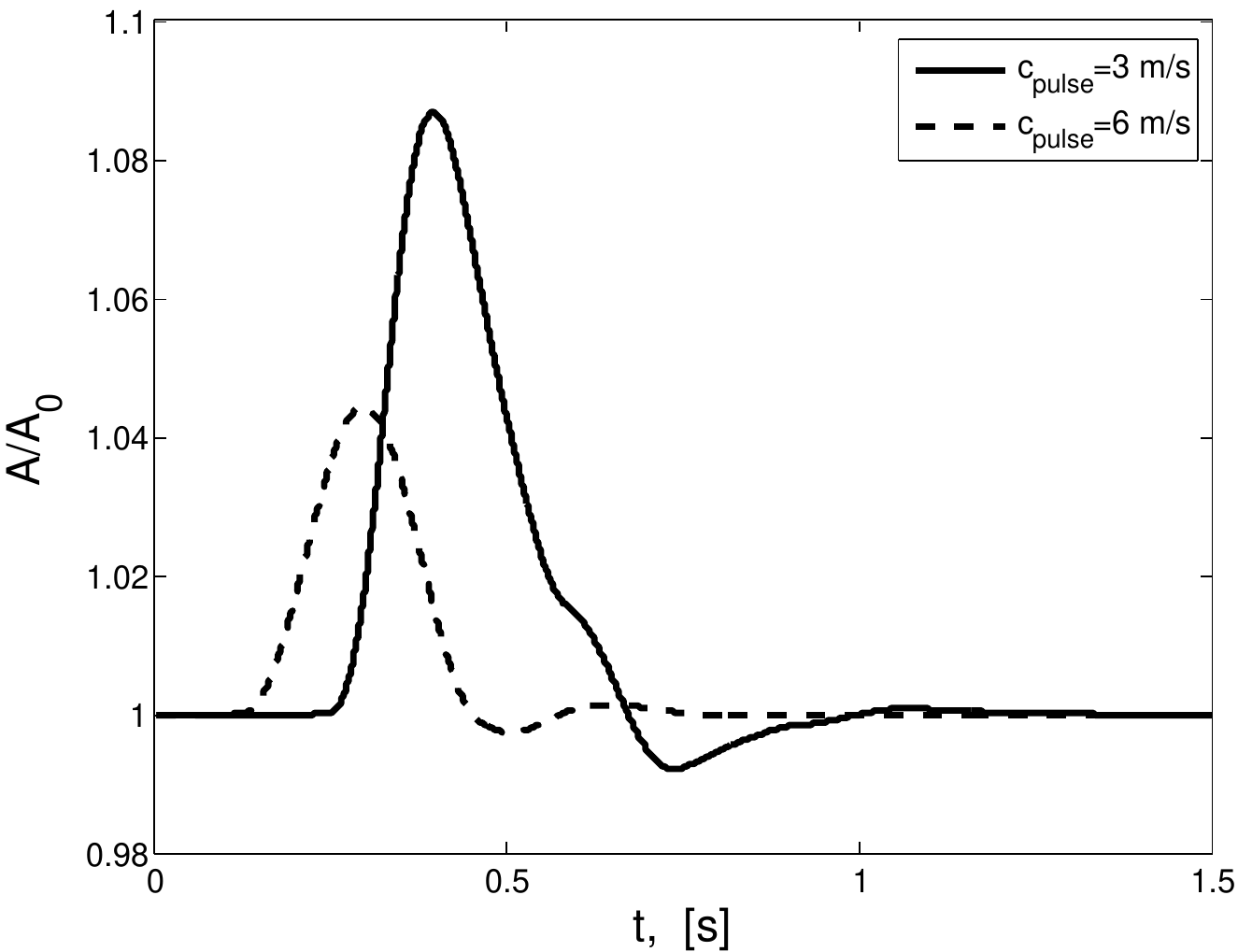}
 \end{minipage}
   \caption{The blood flow velocity (normalized at the initial flow velocity amplitude $u_0=0.25 \, m/s$) and luminal area  profiles (normalized at the undisturbed vessel area $A_0^{(3)}$) measured at the middle of the vessel $3$. For the present case  the  diameters of the vessels 4 and 5 (Fig. 6) are increased and equal $D/\sqrt{3}$, the linear  reflection coefficient for the branching consisting of the vessels $3,4,5$ is now $0.2$. This results in smaller amplitudes of the reflected waves. }
\end{figure}

\subsection{A pulse wave  propagation in a  vessel with prosthesis.}
A semi-infinite vessel with an interior  stiff region  (stent or  prosthesis) is considered.  Similar  test  problem was investigated   using  Galerkin and Taylor-Galerkin finite difference methods in the  papers \cite{2003sherwin1, 2003form_lamp}.

 The mass flow and full energy are continuous across the junctions of the stiff  and  elastic regions \cite{1978lighthill, 2003sherwin}. Then
$$
A_Lu_L=A_Ru_R, \quad  \frac{\rho_0 u^2_L}{2}+p_L=\frac{\rho_0u_L^2}{2}+p_R,
$$
where $L$ and $R$ denote left and  right side  of  the  junction interface respectively and $\rho_0$ is the  blood density. Since the  blood flow  is slow then $p >> \frac{\rho_0 u^2}{2}$ and one can consider 
the pressure continuity property $p_L=p_R$ instead of  the full energy. Consider the Laplace  area-pressure  relation (\ref{p3exp_p}), $n=1/2$.
Using the  linearized pressure pulse wave velocity  $c_{pulse}(A_0)=\sqrt{A_0^{1/2}/\rho_0 D_0}$ the Laplace law reads
$$
p=p_0+2\rho_0 c_{pulse}^2(A_0)\left(\sqrt{\frac{A}{A_0}}-1\right).
$$
The continuity condition at  the junction is considered for the pressure-area relation linearized near $A_0$, i.e. $p=p_0+\rho_0 c_{pulse}^2(A_0)\left(\frac{A}{A_0}-1\right)$. 
In a similar way one can linearize the logarithmic tube law (\ref{p3exp_log}).

In terms  of LB variables the continuity of the mass  flux and the linearized pressure take the following form
$$
(f_{1}-f_{-1})(t, x_n)c_L=(f_{1}-f_{-1})(t, x_{n+1})c_R+\Delta a,
$$
$$
(f_{1}+f_{0}+f_{-1}-A_0)(t, x_n)c_{pulse,L}^2(A_0)=(f_{1}+f_{0}+f_{-1}-A_0)(t, x_n)c_{pulse,R}^2(A_0),
$$
$$
\Delta a \equiv (1/2)(a_R-a_L)\Delta t, 
$$
where the junction is  placed between the lattice  nodes $x_n, x_{n+1}$;  $c_L=\frac{\Delta x_L}{\Delta t}, c_R=\frac{\Delta x_R}{\Delta t}$ are the LB velocities on the left and right sides from the junction; $\Delta x_L, \Delta x_R$  are the LB spatial steps and $\Delta t $ is the LB time step; $a_L, a_R$ are the force  magnitudes on the left and right sides of the junction interface; $c_{pulse,L}^2(A_0)$ and $c_{pulse, R}^2(A_0)$ are the linearized pulse wave velocities. For the  present  problem the lattice  velocities are taken as $c_L^2=3c_{pulse,L}^2(A_0), c_R^2=3c_{pulse,R}^2(A_0)$. Such a calibration is  convenient and guarantees that the lattice sound velocity equals to the linearized pulse wave velocity.

The distribution functions $f_{-1}(t, x_n), f_{1}(t, x_{n+1})$ are  unknowns, they are obtained from the continuity conditions
$$
f_{-1}(t, x_n)=\frac{A_0}{6}+
$$
$$
+\frac{1}{k+k^2}(-k^2\Delta f_{0}(t,x_n)+(k-k^2)\Delta f_{1}(t,x_n)+2\Delta f_{-1}(t,x_{n+1})+\Delta f_{0}(t,x_{n+1})-\Delta a) ,
$$
$$
f_{1}(t, x_{n+1})=\frac{A_0}{6}+
$$
$$
+\frac{1}{1+k}(k^2\Delta f_{0}(t,x_n)+2k^2 \Delta f_{1}(t,x_n)+(k-1)\Delta f_{-1}(t,x_{n+1})-\Delta f_{0}(t,x_{n+1})+\Delta a) ,
$$
where 
$$
\Delta f_{\pm 1}\equiv f_{\pm 1}-\frac{A_0}{6}, \quad k \equiv \frac{c_L}{c_R}=\frac{c_{pulse,L}(A_0)}{c_{pulse,R}(A_0)}.
$$

For the test  problems  considered here  it is assumed   that the  undisturbed vessel cross-sectional area $A_0$ equals $10^{-4}\, m^2$, the blood density $\rho_0$ is $10^3 \, kg/m^3$. 

The first  vessel geometry is depicted in Fig. 2. The  dynamics of a  half-sinusoidal pressure pulse wave with the maximal amplitude of $0.2\, kPa$ and the duration of $0.165 \,s$ starting at the point $x=0$ is studied.  The  pressure-area relation  (\ref{p3exp_log}) is adopted.

The distensibility of  the prosthesis is assumed  to be  $10$ times smaller then in the other part of the vessel, therefore the linearized pulse wave velocity is $\sqrt{10}$ is  larger in the stiffening than in the remaining part of the vessel ( $4 \, m/s$ and $12.6 \, m/s$),  Fig. 2.  

The simulation results show that the pressure maximal magnitude is amplified (in the comparison with the wave at $x=0 \, m$)  near the  left  junction with a subsequent  pressure  drop below the undisturbed level (Fig. 3). This behavior is in a good  agreement with the results from \cite{2003sherwin1, 2003form_lamp}. The pressure increase over the initial pressure values happens due to the  reflection of the wave entering the stiff region while the pressure drop  is observed due to the second  reflection of the wave leaving  the stiff region. 

The second considered  vessel geometry is depicted in Fig. 4. A relatively short vessel segment (Fig. 4) and a small pulse wave amplitude $0.055 \, kPa$ are taken. Such a  choice guarantees that the nonlinear effects  do not distort initial wave shapes and shock waves do not appear. The  distensiblity of the stiff  region is  $100$ smaller than the elastic parts of the vessel, this  means that the  pulse  wave  velocity in the stiff region is $10$ times  larger than in the the elastic segments. The Laplace pressure-area relation is adopted.

The  modeling  results  are  presented  in Fig. 5.   Multiple reflections propagating from the junctions are  clearly visible. The  positive amplitudes correspond  to the reflections occurred  at the  left junction while the waves with negative pressure amplitudes correspond to the reflections at the right junction.
Again, the wave form behavior is very similar  to the results from \cite{2003sherwin1, 2003form_lamp}.

\subsection{Bifurcations of arteries}

Consider a  bifurcation of  $N+1$ vessels:   an undisturbed luminal area  and linearized pulse wave velocity for a parent  vessel are defined as $A_0^{(0)},c_{pulse}(A_0^{(0)})$ and  the same quantities for daughter vessels are  defined as $A_0^{(i)},c_{pulse}(A_0^{(i)}), i=1 \ldots N$. At the junction point  it is assumed that  the continuity  of the pressure  and the conservation of the flow are satisfied \cite{2003sherwin}
\begin{equation}\label{06_bifur01}
\Delta p^{(0)}=\Delta p^{(i)}, i=1\ldots N,
\end{equation}
\begin{equation}\label{06_bifur02}
    A^{(0)}u^{(0)}=\sum_{i=1}^{N} A^{(i)}u^{(i)},
\end{equation}
where $u^{(i)}, \Delta p^{(i)}=p-p_0$ are the blood  velocity and pressure alteration from the diastolic value $p_0$ in the vessels $i=0 \ldots N$.

 For simplicity, the pressure continuity condition  is considered in  the linearized form  for the  pressure-area relation (\ref{p3exp_p}), then one has $\Delta p(A) \equiv p(A)-p_0=\rho_0 c_{pulse}^2(A_0)(A/A_0-1).$ In terms  of the LB variables the equations (\ref{06_bifur01})-(\ref{06_bifur02}) read as
 $$
 (f^{(0)}_{-1}+f^{(0)}_{0}+f^{(0)}_{1}-A_0^{(0)})\frac{c_{pulse}(A_0^{(0)})^2}{A_0^{(0)}}=
 $$
 \begin{equation}\label{06_press_contin}
= (f^{(i)}_{-1}+f^{(i)}_{0}+f^{(i)}_{1}-A_0^{(i)})\frac{c_{pulse}(A_0^{(i)})^2}{A_0^{(i)}},   
\end{equation}
where $i=1\ldots N$ and
\begin{equation}\label{06_flow_contin}
(f^{(0)}_{1}-f^{(0)}_{-1})c^{(0)}= \sum_{i=1}^{N}(f^{(i)}_{1}-f^{(i)}_{-1})c^{(i)}+\Delta a,   
\end{equation}
where $c^{(i)}$ is the lattice step in $i$-s vessel,  moreover
$$
\Delta a=\sum_{i=1}^{N}a^{(i)}\Delta t/2-a^{(0)}\Delta t/2, i=1 \ldots N
$$
where $a^{(i)}, i=0 \ldots N$ is the amplitude of the external force in $i$-s vessel (the definition of the external virtual force is given in (\ref{force_definition})). The time step $\Delta t$ is constant in all vessels, while the lattice velocities  $c^{(i)}=\Delta x^{(i)}/\Delta t$ can be different, here $\Delta x^{(i)}$ is the lattice spatial step. The  lattice  velocities $c^{(i)}$ are taken in a such way  that
$$
\frac{c^{(i)}}{c^{(j)}}=\frac{c_{pulse}(A_0^{(i)})} { c_{pulse}(A_0^{(j)}) },
$$
the latter property guarantees that the pulse wave propagation speed will be different  if $c_{pulse}(A_0^{(i)})$ are non-equal in the vessels.

At the  junction point  $x$ the value of  $f_{-1}^{(0)}(t,x)$ is unknown for the parent vessel and the values of $f_{1}^{(i)}(t,x)$ are unknown for the daughter vessels $i 1 \ldots N$, they  should be  found from the equations 
(\ref{06_press_contin})-(\ref{06_flow_contin}). One has
\begin{equation}\label{06_sol1}
f_{-1}^{(0)}=f_{1}^{(0)}-\sum_{i=1}^{N}(f^{(i)}_{1}-f^{(i)}_{-1})k^{(i)}+\Delta b,
\end{equation}
where $f^{(i)}_{1}, i=1 \ldots N$ are calculated from the linear system 
$$
\sum_{j \neq i}^{N}k^{(j)}\Delta f_1^{(j)}+
(k^{(i)}+(r^{(i)})^2)\Delta f_1^{(i)}=
$$
\begin{equation}\label{06_sol2}
\sum_{j=1}^N k^{(j)} \Delta f_{-1}^{(j)} -(r^{(i)})^2(\Delta f_{0}^{(i)} +\Delta f_{-1}^{(i)} )+(2\Delta f_{1}^{(0)} +\Delta f_{0}^{(0)} )-\Delta b,
\end{equation}
here $i=1 \ldots N$,
and the following notations are introduced
$$
\Delta b \equiv\frac{\Delta a}{c^{(0)}} , \quad k^{(i)} \equiv\frac{c^{(i)}}{c^{(0)}}  , \quad r^{(i)} \equiv k^{(i)}\sqrt{\frac{A^{(0)}}{A^{(i)}}}, \quad \Delta f_{k}^{(i)}=f_{k}^{(i)}-w_kA_0^{(i)}.
$$
It is assumed that the wave  is absorbed by the vessel boundaries (the vessels are well-matched with the distal vasculature). In order to absorb the incident   wave the impedance boundary conditions should be  stated \cite{2013schlaffer}.

The modeling artery geometry is depicted in Fig. 6.  The input velocity wave  is prescribed to the left inlet of the vessel $1$ and has the  following form $0.25*sin(\pi t/0.33) \,\, m/s$ for $0 \leq t \leq 0.33 \, s$. 
The  linearized pulse wave velocities are taken same  in all vessels: in the  first case  the value $3 \, m/s$ is  used, in the second case this velocity equals $6 \, m/s$. The  case  of the increased pulse  velocity corresponds to the more stiff vessels which results to the smaller amplitudes of luminal area changes (Fig. 7-8). The Laplace pressure-area relation is adopted for the present problem.
For the first (left) bifurcation the linear reflection coefficient  \cite{1978lighthill} is
$$
R=\frac{\frac{A_0^{(0)}}{c_0^{0}}-\frac{A_0^{(1)}}{c_0^{1}}-\frac{A_0^{(2)}}{c_0^{2}}}{\frac{A_0^{(0)}}{c_0^{0}}+\frac{A_0^{(1)}}{c_0^{1}}+\frac{A_0^{(2)}}{c_0^{2}}}
$$
and equals $0$ for the forward wave traveling to the  junction from the the vessel $1$, on the  other hand,  for the backward wave  traveling to this  junction from the  vessel $2$  or $3$ this coefficient  equals $-0.5$.
The  linear reflection coefficient at the second (right) junction equals $0.5$ for the forward wave in the vessel $3$. The backward waves at the distal ends of the vessels $4$ and $5$ do not exist since the outlets of these vessels absorb incident waves. Thus, the wave  is  partially "trapped" in the  vessel 3. 

The wave dynamics and multiple wave reflections are measured at the middle of the vessel $3$, see Fig. 7-8. In Fig. 8 the modeling results are presented for the geometry in Fig. 6 where the vessels 4 and 5 have increased radius ($D/\sqrt{3}$). This results in smaller reflection coefficient $0.2$ and  therefore smaller amplitudes of the reflected waves in comparison with the profiles in Fig. 7.

The  problem with the geometry in Fig. 6 was also  considered in the paper \cite{2003sherwin}. The  presented  area and velocity profiles  are very similar  to the data from \cite{2003sherwin}.

\begin{figure}
   \centering
  \begin{minipage}[t]{.99\textwidth}
       \includegraphics[width=0.95\textwidth]{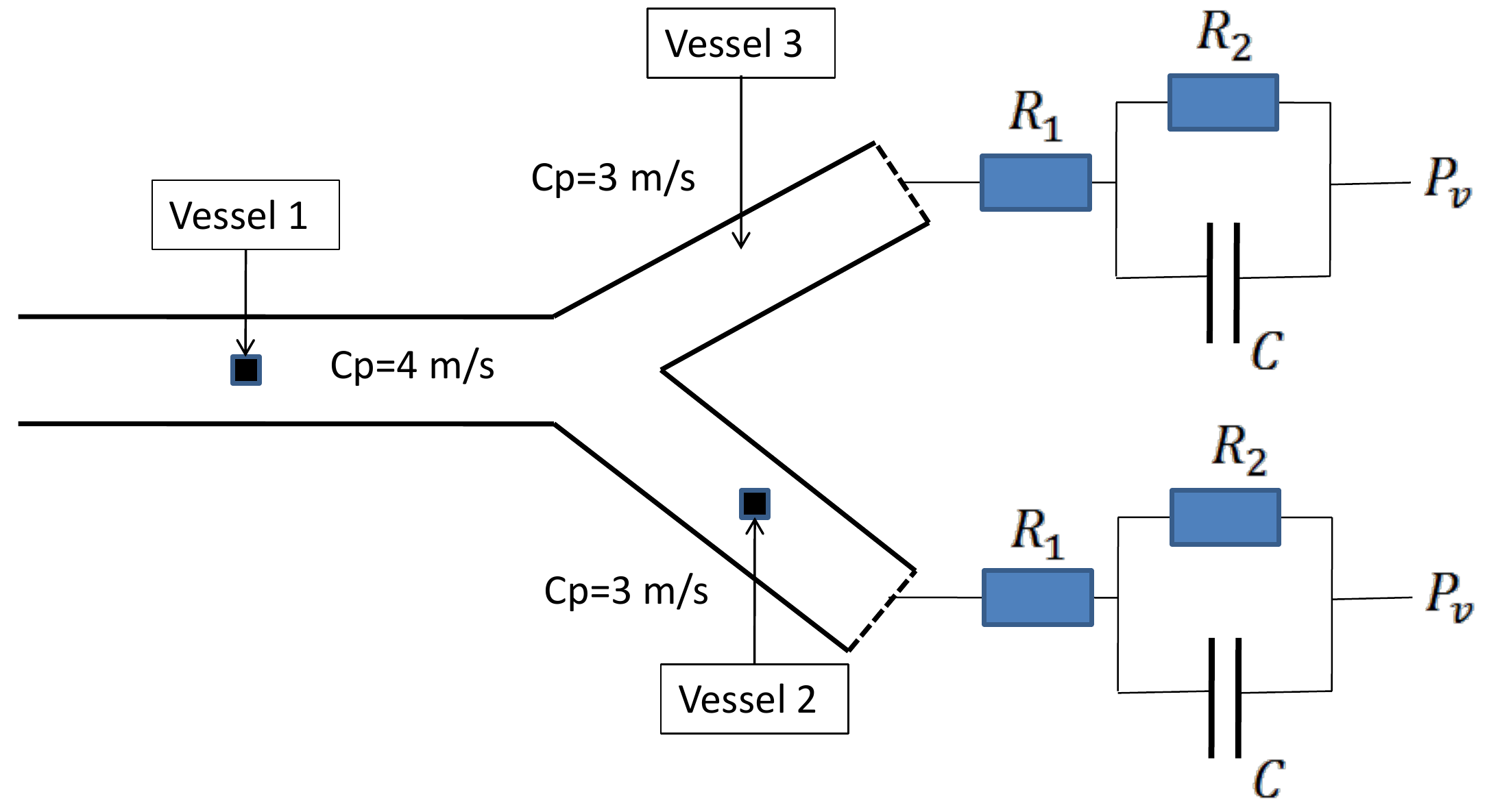}
 \end{minipage}
   \caption{ Modeled vessel branching geometry with lumped elements models (RCR models). The  parent vessel (vessel 1) having the length $0.1 \, m$ and  the linearized pulse wave velocity  $4 \, m/s$ is connected with two daughter vessels (vessel 2 and vessel 3) having the
   length $0.05 \, m$ and the linearized pulse wave velocity $3 \, m/s$.
   The daughter vessels (vessel 2 and vessel 3) are connected with RCR models which are responsible for  microcirculation, $R_1$ is the characteristic impedance of the daughter vessels, $R_2, C$ are the  resistance and compliance  of the microcirculation, $P_v$ is the venous pressure (equals zero in the  present case).
     }
\end{figure}

\begin{figure}
   \centering
  \begin{minipage}[t]{.99\textwidth}
       \includegraphics[width=0.9\textwidth]{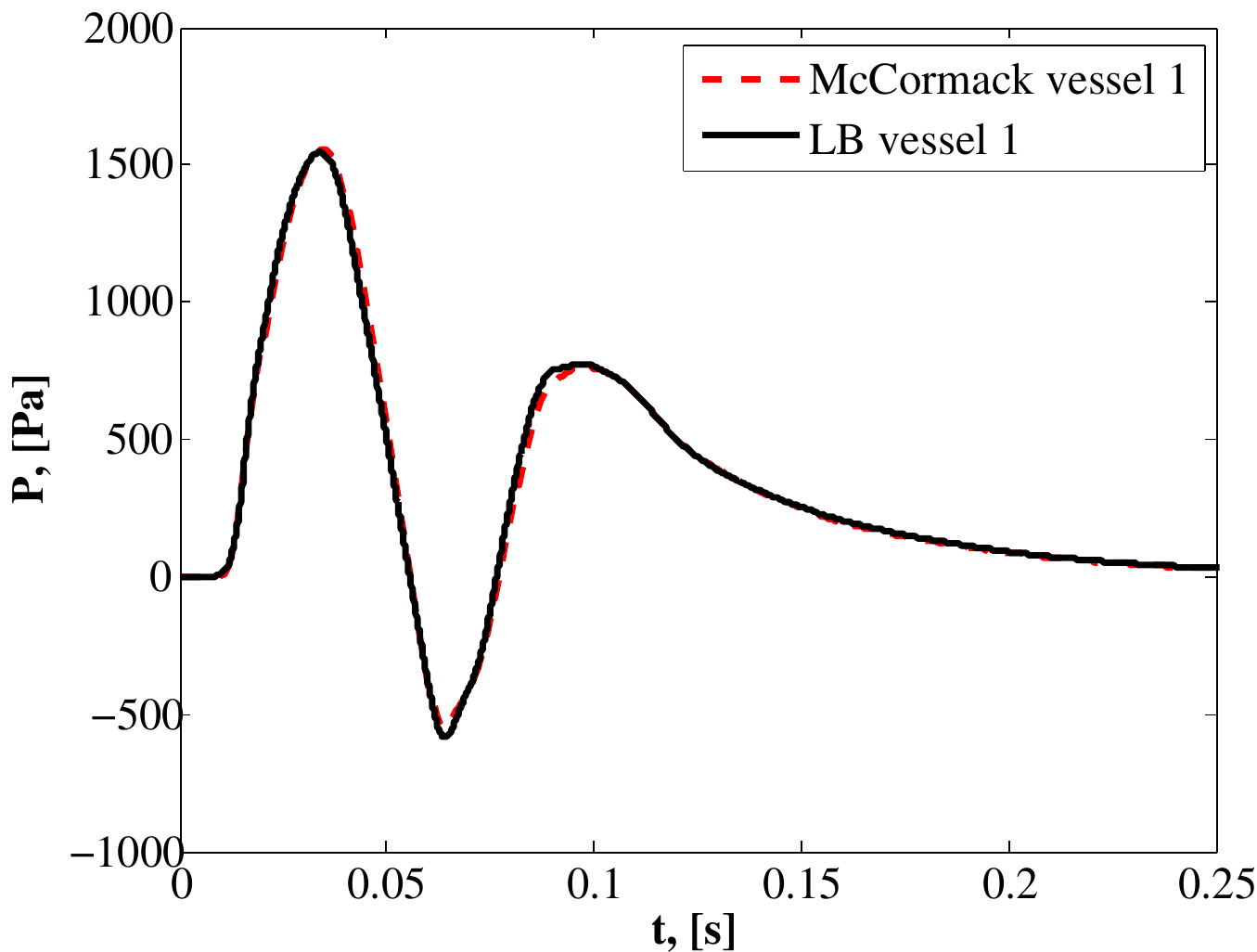}\\
       \includegraphics[width=0.9\textwidth]{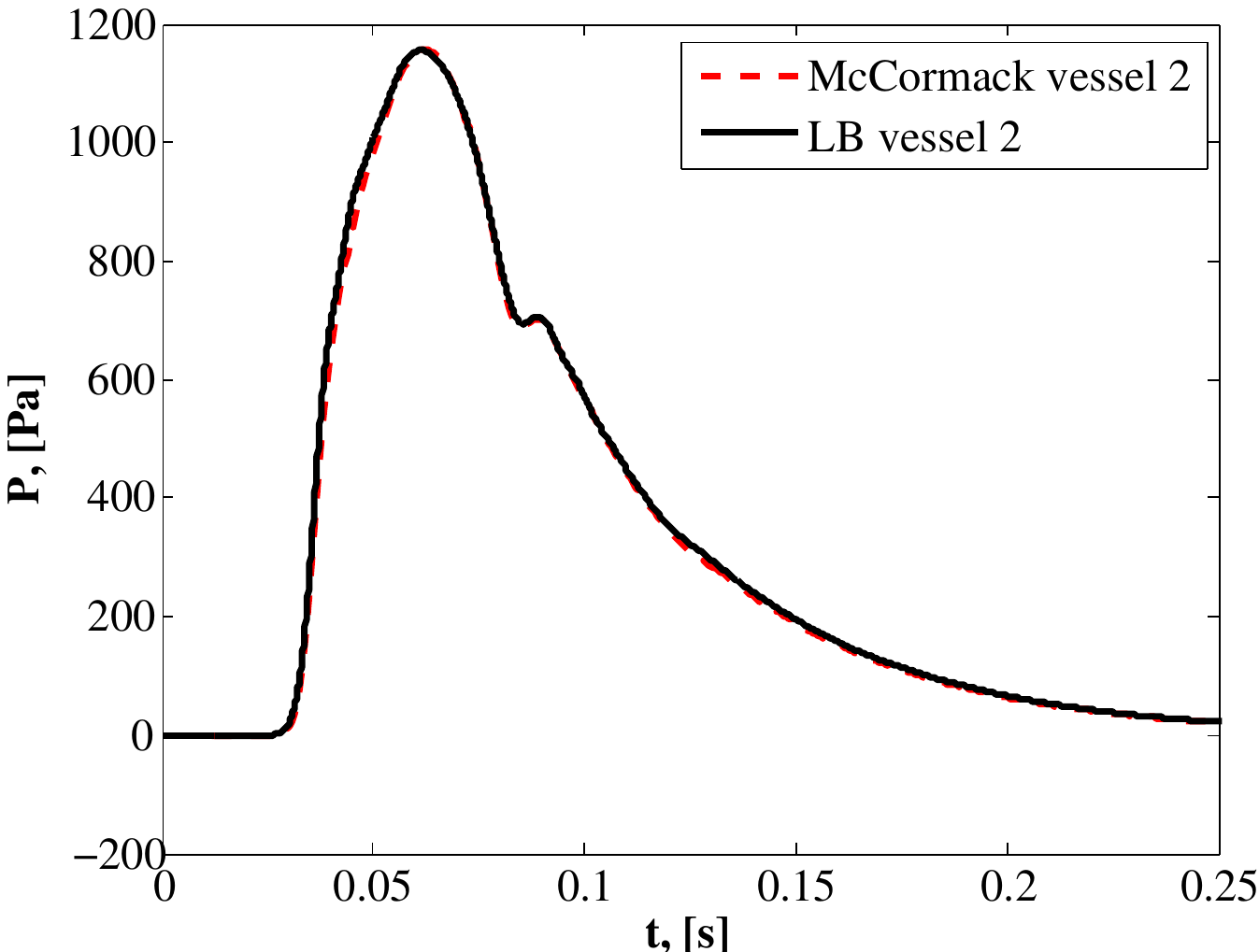}
 \end{minipage}
   \caption{ Pressure profiles obtained using McCormack difference scheme (dashed line) and LB method (solid line)  at the  middle of the vessel 1 (upper figure) and vessel 2 (lower figure) for the vessel geometry depicted in Fig. 9.
     }
\end{figure}
\begin{figure}
   \centering
  \begin{minipage}[t]{.99\textwidth}
       \includegraphics[width=0.9\textwidth]{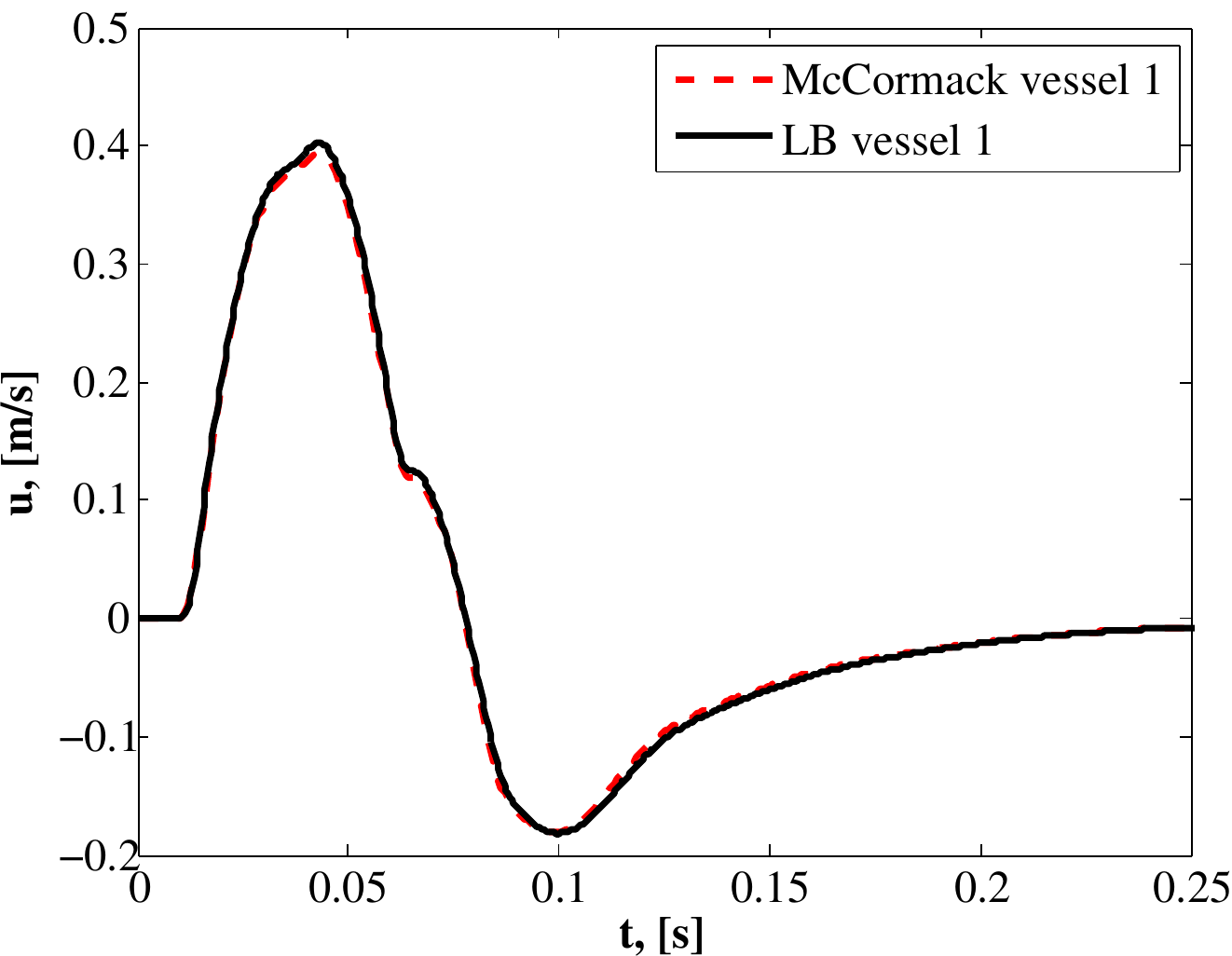}\\
       \includegraphics[width=0.9\textwidth]{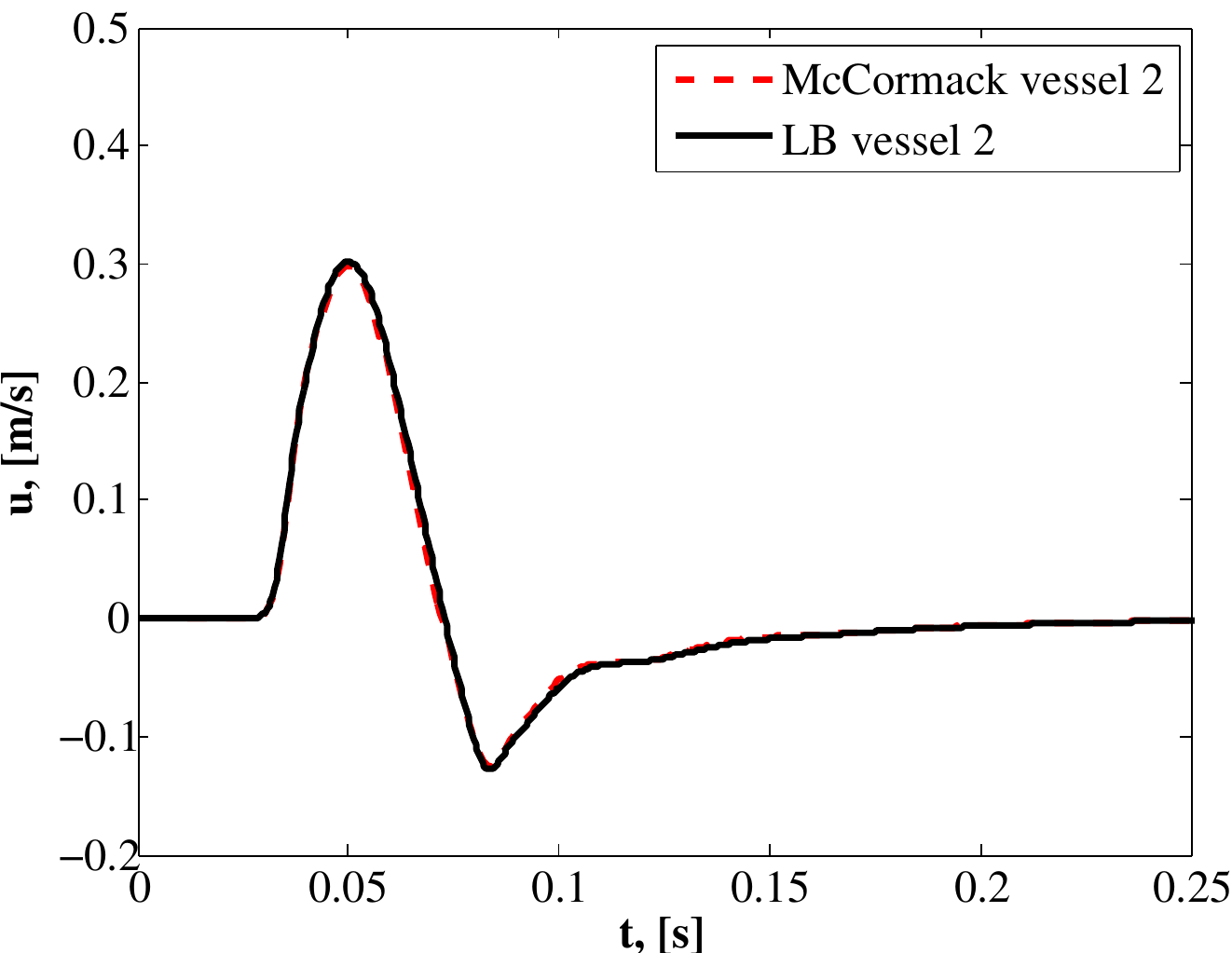}
 \end{minipage}
   \caption{ Velocity profiles obtained using McCormack difference scheme (dashed line) and LB method (solid line)  at the  middle of the vessel 1 (upper figure) and vessel 2 (lower figure) for the vessel geometry depicted in Fig. 9.
     }
\end{figure}

\subsection{RCR boundary conditions}

Resistor-Capacitor-Resistor (RCR)  boundary conditions \cite{1971westerhof, 2008alastruey, 2011shi} are very popular in hemodynamical modeling. These  boundary  conditions are analogous  to electric  circuit with two resistances $R_1, R_2$ and one capacitor $C$ (Fig. 9). The  first resistance $R_1$  accounts for the distal part  of the truncated vessel and  is usually selected in such a way that there are no wave reflection in a junction between a vessel and RCR model, while  the second resistor $R_2$ accounts for the  drag  forces in the distal arterioles, the capacitor $C$  term is responsible for the compliance of the arterioles. One has  for the pressures and velocities $p,q$ at the distal ends  of the vessels 2,3 the following equation
$$
C\frac{dp}{dt}+\frac{p}{R_2}= CR_1\frac{dq}{dt}+\left(1+\frac{R_1}{R_2}\right)q,
$$
the first order discretization of this equation is following
$$
(R\Delta t+R_1R_2C)q(t+\Delta t)=(\Delta t +R_2 C)p(t+\Delta t)-R_2C(p(t)-R_1q(t)),
$$
where $R=R_1+R_2$ and $\Delta t$ is the  time  step for the  difference scheme. It is assumed that the values $p(t),q(t)$ are known.

Any traveling wave in a vessel  can be separated into the   forward and backward  components  by the rule
\begin{equation}\label{eq05_separ}
p_f=\frac{1}{2}(p+R_1q ), \quad  p_b=\frac{1}{2}(p-R_1q ).
\end{equation}
Using  (\ref{eq05_separ})
one  has
$$
(R\Delta t+R_1R_2C)(p_f(t+\Delta t)-p_b(t+\Delta t))=
$$
$$
=R_1(\Delta t +R_2 C)(p_f(t+\Delta t)+p_b(t+\Delta t))-R_2CR_1(p(t)-R_1q(t))
$$
then
\begin{equation}\label{eq05_pb}
p_b(t+\Delta t)=
\frac{R_2\Delta t p_f(t+\Delta t) +R_1R_2C   (p(t)-R_1q(t))}{(R+R_1)\Delta t+2R_1R_2C}.
\end{equation}
Next, we mention that 
\begin{equation}\label{eq05_pf}
p_f(t+\Delta t)\equiv p_f(t+\Delta t, L)=p_f(t, L-c_{pulse}\Delta t)
\end{equation} 
since $p_f$ can be considered  as the  forward traveling wave  with the  velocity $c_{pulse}$, where $x=L$ is the  distal point  of the considered vessel. Therefore, we  have  expressed $p_f(t+\Delta t, L), p_b(t+\Delta t, L)$  via known functions
$p_f(t, L), p_b(t, L), p_f(t, x-c_{pulse}\Delta t)$.
Then
\begin{equation}\label{eq05_sum_p}
p(t+\Delta t, L)=(p_f(t+\Delta t, L)+p_b(t+\Delta t, L)),
\end{equation}
\begin{equation}\label{eq05_sum_q}
q(t+\Delta t, L)=R_1^{-1}(q_f(t+\Delta t, L)+q_b(t+\Delta t, L)).
\end{equation}
Now one expresses $p,q$ via the lattice Boltzmann variables 
$$
\Delta f_{-1}+\Delta f_{0}+ \Delta f_{1}=C_0p(t+\Delta t,L),
$$
$$
(\Delta f_{1}+ \Delta f_{-1})c+ \frac{\Delta t a(L)}{2}=q(t+\Delta t,L)
$$
where $a$ is the amplitude  of the external force,  $C_0=\frac{A^{(0)}}{\rho c_{pulse}^2}$, where $c_{pulse}, A^{(0)}$ are the linearized pulse velocity and  undisturbed luminal area, $a(L)$ is approximated as $2a(L-1)-a(L-2)$. The values $f_0(t+\Delta t, L), f_{-1}(t+\Delta t, L)$ are unknowns at the distal end of the vessel, they are found from the equations 
\begin{equation}\label{eq10}
f_{-1}(t+\Delta t, L)=\frac{1}{6}A^{(0)}+\Delta  f_{1} +\frac{\Delta t a(L)}{2}-\frac{q(t+\Delta t,L)}{c},  
\end{equation}
\begin{equation}\label{eq11}
f_{0}(t+\Delta t, L)=\frac{4}{6}A^{(0)}-2\Delta  f_{1} +C_0p(t+\Delta t,L)-\frac{\Delta t a(L)}{2}+\frac{q(t+\Delta t,L)}{c}.  
\end{equation}
The relations (\ref{eq10})-(\ref{eq11}) define RCR boundary conditions for the LB model.  Finally one can see that the  unknown values of LB distribution functions at the  right end  of the  vessel (distal end) $x=L$ are  defined by $p(t+\Delta t,L), q(t+\Delta t,L)$, they can be represented as a sum of  the forward and  backward  components (\ref{eq05_sum_p})-(\ref{eq05_sum_q})  which in turn depend on $p_f(t, L), p_b(t, L), p_f(t, x-c_{pulse}\Delta t)$ (the formulas (\ref{eq05_pb}), (\ref{eq05_pf})).

As a test  problem the propagation of  a pulse wave in a vessel network consisting of three vessels (one parent vessel and two daughter vessels) and one bifurcation connected with RCR models at the distal outlets of the daughter vessels is considered.  The  parent vessel has the  length $0.1 \, m$ , the luminal radius equals $0.01 \, m$, the linearized pulse wave  velocity equals $4 \, m/s$.  For the daughter vessels the values of the vessel length, luminal radius, linearized pulse wave  velocity are taken as
$0.05\, m, 0.009\, m,  3 \, m/s$ respectively.
The  Laplace law is adopted for the pressure-area relation. The initial pressure profile has the duration of $0.05 \, s$ and the  maximal amplitude corresponding to $10 \%$ luminal area increase in the vessel 1. The value of $R_1$  equals $\rho_0 c_{pulse}(A^{(0)})/ A^{(0)}$, where $c_{pulse}(A^{(0)}), A^{(0)}$ are the linearized pulse velocity and  undisturbed luminal area respectively. The full resistance $R_1+R_2$ and compliance $C$ are taken as $10^8 Pa \cdot s/m^3, 10^{-9} \, m^3/Pa$ respectively, then the  diastolic decay time $\tau \equiv R_2C \approx 0.09 \, s $.

The choice of the modeling parameters results in relatively complicated behavior of the pressure wave: since the linear reflection coefficient is negative for the vessel 1 at the branching point ($\approx -0.367$) the negative  pressure wave component appears with subsequent   positive  backward wave caused by the microcirculation (Fig. 10, upper figure); the pressure component in the vessel 2 contains the forward pressure wave passed through the bifuracation and the backward pressure wave caused by the microcirculation (Fig. 10, lower figure).

As a benchmark solution the modeling results for the second order McCormack difference scheme \cite{2007hirsch} are adopted. For the McCormack scheme at the bifurcation point the continuity equations for the flux and  pressure are considered jointly with the requirement that the characteristic variables are constant \cite{2003sherwin}. The realization of the boundary conditions at the vessel junction for the McCormack scheme differs from the LB approach (\ref{06_sol1})-(\ref{06_sol2}) in which the characteristics are not considered.

The  pressure and velocity profiles measured at  the middle  of the  vessels $1,2$ are depicted in Fig.10 and Fig.11 for LB model and McCormack difference scheme. Both methods  give  very similar  profiles, small differences are observed  at some  points, they can be attributed to the different approximation methods of the  boundary conditions at the vessel bifurcations.

\section{Conclusion}
In the  present  paper the  method for  modeling of the  1D blood flow equations   in elastic vessels  using kinetic LB approach is  proposed. The  inclusion of the virtual external force in LB model  allows  to mimic arbitrary vessel luminal area response  to the exerted blood pressure. Several test problems are  considered.

In comparison to the conventional numerical methods of 1D blood flow modeling LB approach has the following  distinctive  features.
    1.  In LB approach the  viscosity  of the  fluid is  the inherent property and its value is linearly dependent on $\tau$. This  property can be valuable  for the  modeling  of lengthy vessel networks in which the  diffusion effects are non-negligible.
    2.  The implementation of the  boundary  conditions for junctions presented in the paper differs  from the conventional realization. In the  presented  method the  characteristics are not considered, the conditions for the pressure and the velocity in the junction are written explicitly in terms of LB variables. It is  also possible to implement these boundary conditions  in a standard way by expressing the characteristics in terms  of LB variables (in a similar way  like RCR conditions are introduced).
    The modeling results  of LB approach are very similar (not exactly the same) to the  pressure  and velocity profiles obtained  with use  of the MacCormack scheme with the application of the conventional boundary conditions at the  junctions. 
    3.  The  abrupt changes  in materials  properties  are allowed in the presented approach. Note that for some  discrete methods  the discontinuities in elastic properties are inadmissible \cite{2003form_lamp}.
    On the  other  hand the  inclusion  of the vessel parts in which the pulse wave velocity changes smoothly  over a some spatial region is  not trivial for the  presented  method. 4. In LB approach the advection is linear (streaming part is linear) while for the conventional methods the discretization of the nonlinear terms $u\frac{\partial \rho}{\partial x}, u\frac{\partial u}{\partial x}$ should be  considered. The linearity of the advection part is an  attractive property  for the  parallel multi-CPU implementation of LB method.

The  implementation of viscoelastic vessel response was not considered  in the  present paper.  Potentially  the viscoelastic   effect  can be inserted in the virtual force term. This  term will contain an approximation of the vessel area time derivative (Kelvin-Voigt model), therefore an additional theoretical question about the stability of the scheme should be considered. This question is  leaved  for the future study.

\bibliography{main}

\end{document}